\documentclass[%
 aip,
 amsmath,amssymb,
 reprint,
 floatfix]{revtex4-2}

\usepackage{amsthm}
\usepackage{algorithm}
\usepackage{appendix}
\usepackage{graphicx}
\usepackage{dcolumn}
\usepackage{bm}
\usepackage[utf8]{inputenc}
\usepackage{mathptmx}
\usepackage{multirow}
\usepackage{braket}
\usepackage[english]{babel}
\usepackage{xcolor}
\colorlet{RED}{red}
\colorlet{BLUE}{blue}
\usepackage{dcolumn}
\usepackage{bm}
\usepackage[version=4]{mhchem} 
\usepackage{acronym}
\usepackage{adjustbox}
\usepackage{float}
\usepackage{microtype} 

\usepackage[margin=2.3cm,bmargin=1cm,footnotesep=1cm]{geometry}

\usepackage{tikz}
\usetikzlibrary{automata,positioning}
\usetikzlibrary{shapes.geometric, arrows}
\tikzstyle{startstop} = [rectangle, rounded corners, text centered, draw=black, fill=none]
\tikzstyle{io} = [trapezium, trapezium left angle=70, trapezium right angle=110, text centered, draw=black, fill=blue!30]
\tikzstyle{process} = [rectangle, text centered, draw=black, fill=none]
\tikzstyle{decision} = [diamond, aspect=1.5, text centered, draw=black, fill=none]
\tikzstyle{arrow} = [thick,->,>=stealth]

\usepackage{physics}
\usepackage{bm}
\usepackage{xcolor}
\usepackage{wrapfig}
\usepackage{float}

\begin{document}

\title{Exploring the exact limits of the real-time equation-of-motion coupled cluster cumulant Green's functions}
\author{Bo Peng}
\email{peng398@pnnl.gov}
\affiliation{%
Physical and Computational Science Directorate, Pacific Northwest National Laboratory, Richland, Washington, 99354, USA
}

\author{Himadri Pathak}
\affiliation{Advanced Computing, Mathematics, and Data Division, Pacific Northwest National Laboratory, Richland, Washington 99354, USA}

\author{Ajay Panyala}
\affiliation{Advanced Computing, Mathematics, and Data Division, Pacific Northwest National Laboratory, Richland, Washington 99354, USA}

\author{Fernando D. Vila}
\affiliation{Department of Physics, University of Washington, Seattle, WA 98195}

\author{John J. Rehr}
\affiliation{Department of Physics, University of Washington, Seattle, WA 98195}

\author{Karol Kowalski}
\affiliation{%
Physical and Computational Science Directorate, Pacific Northwest National Laboratory, Richland, Washington, 99354, USA
}
 
\date{\today}

\begin{abstract}
In this paper, we analyze the properties of the recently proposed real-time equation-of-motion coupled-cluster (RT-EOM-CC) cumulant Green’s function approach [J. Chem. Phys. 2020, 152, 174113]. We specifically focus on identifying the limitations of the original time-dependent coupled cluster (TDCC) ansatz and propose an enhanced double TDCC ansatz ensuring the exactness in the expansion limit. Additionally, we introduce a practical cluster-analysis-based approach for characterizing the peaks in the computed spectral function from the RT-EOM-CC cumulant Green’s function approach, which is particularly useful for the assignments of satellite peaks when many-body effects dominate the spectra. Our preliminary numerical tests focus on reproducing, approximating, and characterizing the exact impurity Green's function of the three-site and four-site single impurity Anderson models using the RT-EOM-CC cumulant Green’s function approach. The numerical tests allow us to have a direct comparison between the RT-EOM-CC cumulant Green's function approach and other Green's function approaches in the numerical exact limit.
\end{abstract}

\maketitle

\section{Introduction}

Attosecond laser pulses exhibit a broad spectral range and relatively high intensity, pioneering ultrafast research such as delayed photoemission,\cite{doi:10.1126/science.aao7043,doi:10.1126/science.1189401,PhysRevLett.117.093001} electronic response to sudden ionization,\cite{Goulielmakis2010,doi:10.1126/science.aab2160,doi:10.1126/science.adn6059} charge localization and transfer in molecules,\cite{Sansone2010,doi:10.1126/science.1254061} autoionization absorption,\cite{doi:10.1126/science.1234407} and conductivity control in dielectrics,\cite{Schultze2013} to name a few. Typically, in instantaneous processes lasting up to a few femtoseconds, the electronic dynamics can be considered free from nuclear motion, allowing theoretical treatments to focus solely on solving the time-dependent electronic Schrödinger equation within the Born-Oppenheimer approximation, which directly corresponds to experimental setups. Various time-dependent electronic structure methods have been developed in both the frequency and time domains. For example, real-time time-dependent density-functional theory~\cite{doi:10.1021/acs.chemrev.0c00223}, \textcolor{black}{including its extensions treating the relativistic effects\cite{Repisky2015_relativistic} and above-ionization core-level excitations,\cite{Lopata2015_Xray}} often demonstrates a reasonable balance between computational efficiency and accuracy. Nevertheless, for dynamics of the electronic excited states that feature strong electron correlation and often involve multiple configurations, multi-configurational time-dependent Hartree-Fock methods~\cite{MEYER199073,BECK20001,Meyer2003} or active space self-consistent field methods~\cite{PhysRevA.88.023402,PhysRevA.87.062511,PhysRevA.90.062508,PhysRevA.91.023417,10.1063/1.5020633} can be employed for higher accuracy, albeit with very demanding computational costs. Alternative high-level approaches that exhibit modest polynomial scaling with the capability of systematic improvement in accuracy usually focus on time-dependent coupled cluster (TDCC) theory.\cite{https://doi.org/10.1002/qua.560120850,PhysRevA.28.1217,https://doi.org/10.1002/wcms.1666} Previous efforts have demonstrated the capability of the TDCC formulation in identifying excitation energy,\cite{10.1063/1.451241} \textcolor{black}{(including computing core-excitation spectra\cite{Bartlett2019_core})} computing linear response properties,\cite{PhysRevA.28.1217} computing spectral functions\cite{PhysRevB.18.6606} and linear absorption spectra in ultraviolet and X-ray energy regimes,\cite{doi:10.1021/acs.jctc.6b00796,10.1063/1.5125494,doi:10.1021/acs.jpclett.7b01206,PhysRevA.102.023115} including the incorporation of relativistic wave functions,\cite{doi:10.1021/acs.jctc.9b00729} finite temperature and non-equilibrium formalism,\cite{doi:10.1021/acs.jctc.8b00773,doi:10.1021/acs.jctc.9b00750} reduced scaling schemes,\cite{doi:10.1021/acs.jpca.3c05151} and adaptive numerical integration.\cite{doi:10.1021/acs.jctc.2c00490} Remarkably, besides electronic dynamics, TDCC theory also applies to nuclear dynamics\cite{PhysRevC.18.2380,PhysRevC.19.1971,PhysRevC.86.014308} and vibrational states or dynamics.\cite{10.1063/5.0190034,10.1063/5.0186000,10.1063/1.472170,10.1063/5.0015413}

On the other hand, the Green's function (GF) approach\cite{Hedin99review} is often employed to treat electron correlation in excited electronic states. Electron correlation is crucial for determining and characterizing the quasi-particle (QP) and satellite peaks observed in, for example, X-ray photoemission spectra (XPS). To effectively capture this correlation in excited states, various theoretical approaches have incorporated the Green's function formulation. These include perturbative treatments,\cite{10.1063/1.4884951,10.1063/1.4994837,10.1063/1.1884965} algebraic diagrammatic construction,\cite{PhysRevA.26.2395,PhysRevA.28.1237,https://doi.org/10.1002/wcms.1206,10.1063/1.5131771} dynamical mean-field theory,\cite{RevModPhys.68.13,RevModPhys.78.865} GW approximations,\cite{hedin65_a796,FAryasetiawan_1998,PhysRevLett.96.226402,PhysRevB.89.155417} and ground state coupled cluster theory.\cite{nooijen92_55, nooijen93_15, nooijen95_1681,PhysRevB.93.235139,kowalski18_4335,berkelbach18_4224,PhysRevB.100.115154,shee2019CC}

Combining TDCC theory with Green's function formalism, Schönhammer and Gunnarsson have demonstrated the computation of the core-hole Green's function from the phase factors of the TDCC wave function.\cite{PhysRevB.18.6606} Building on their formulation, we have recently developed a real-time equation-of-motion coupled cluster (RT-EOM-CC) cumulant GF approach,\cite{rehr2020equation,vila2020real, vila2021equation, vila2022real, vila2022_water} in which the Green's function formulation adopts an exponential cumulant form to build up correlation in excited states, analogous to ground state coupled cluster theory. Technically, the cumulant is obtained by solving coupled ordinary differential equations of the TDCC amplitudes, providing higher-order vertex corrections to the one-particle self-energy compared to the traditional cumulant approximation\cite{10.1063/1.4934965,LarsHedin_1999,PhysRevB.93.235139} and the stochastic vertex approximation.\cite{doi:10.1063/5.0044060} Numerical results have shown the applications of our RT-EOM-CC cumulant GF approach in reproducing the XPS of small-to-medium molecules described by moderate basis sets.\cite{vila2021equation, vila2022real, vila2022_water} With heterogeneous parallel implementation and tensor algebraic techniques, RT-EOM-CC simulations of over 2,000 spin-orbitals have been achieved.\cite{doi:10.1021/acs.jctc.3c00045}

While advancing towards larger-scale RT-EOM-CC simulations to resolve the many-body effects in the spectroscopy of more complex molecular systems, another fundamental aspect to consider is the exactness of the introduced TDCC ansatz in the computation of Green's functions and its alignment with the actual many-body physical picture of electron transitions. Our previous RT-EOM-CC results in computing spectral functions associated with QP and satellites, when compared with other theoretical approaches, show great agreement with experimental results. For instance, within the single and double excitation manifold, RT-EOM-CC results for some small molecules seemed more accurate than those obtained with CCGF using the modest basis set.\cite{vila2021equation, vila2022real, vila2022_water} However, in weakly correlated scenarios within a single reference framework, the one-particle Green's function computed by the CCGF approach can be exact in the expansion limit. \textcolor{black}{Even though the original TDCC ansatz in the RT-EOM-CC approach\cite{PhysRevB.18.6606} theoretically allows for ``exact'' $(N-1)$-particle dynamics at the full $(N-1)$-particle expansion limit, the ansatz didn't explicitly account for the influence of correlations in the $N$-particle space. Therefore, it remains unclear whether the exact one-particle Green's function, which corresponds to scenarios involving changes in particle number, can be achieved using this same ansatz in the RT-EOM-CC framework}. In this paper, by explicitly considering both the $N$ and $(N-1)$-particle spaces in the RT-EOM-CC cumulant GF approach, we examine the quality of the previous TDCC ansatz, and propose an enhanced new TDCC ansatz and its approximations that features the double CC formulation and are capable of incorporating different Fock spaces without modifying the Hamiltonian. We then analyze the impact of different ans\"{a}tze on the computed Green's functions. Moreover, we propose a scheme for addressing the component analysis of the Green's function computed by the RT-EOM-CC approach, which provides a powerful tool for characterization and peak assignment of the computed spectral functions. Our preliminary numerical test focuses on the single-impurity Anderson model (SIAM) with a limited number of bath sites, where high-level theoretical results and exact solutions can be obtained to test the performance and determine the exact limits of our proposed TDCC ans\"{a}tze in RT-EOM-CC simulations.


\section{Theory}

\subsection{One-particle Green's function}
Given an electronic Hamiltonian $H$ for an $N$-electron system with the ground state $| \Psi^{(N)}\rangle$ and the corresponding energy $E_g^{(N)}$, the (retarded) one-particle Green's function at an occupied spin-orbital $c$ can be expressed as
\begin{align}
    G_c^{\rm{R}}(t) 
    &= -i \Theta(t) \langle \Psi^{(N)} | \big[ a_c(t), a_c^\dagger(0) \big]_+ | \Psi^{(N)}\rangle \notag \\
    &\approx -i \Theta(t) \exp(-iE_g^{(N)} t) \times \notag \\
    &~~~~~~~~ \langle \Psi_c^{(N-1)} | \exp(iHt) | \Psi_c^{(N-1)}\rangle.  \label{eq: chGF}
\end{align}
Here, we assume $a_c^\dagger | \Psi^{(N)}\rangle \approx 0$ because of the occupation of spin-orbital $c$ in the $N$-electron ground state. $\Theta(t)$ is the Heavyside step function ensuring causality. $| \Psi_c^{(N-1)}\rangle = a_c | \Psi^{(N)}\rangle$ denotes the (non-equilibrium) $(N-1)$-electron state generated by removing one electron at the spin-orbital $|c\rangle$ from the $N$-electron ground state wave function $| \Psi^{(N)} \rangle$. $a_c(t)$ and $a_c^\dagger(t)$ are the annihilation and creation operators at time $t$ in the Heisenberg picture, i.e.
\begin{align}
a_c(t) &= \exp(iHt)a_c \exp(-iHt),\\
a_c^\dagger(t) &= \exp(iHt)a_c^\dagger \exp(-iHt).
\end{align}
%

Employing the bi-orthogonal coupled cluster (CC) wave function ans\"{a}tze~\cite{arponen83_311,Szalay1995ECC, PIECUCH1999XCC,salter1987property,stanton93_7029,monkhorst77_421,jorgensen90_3333,nooijen92_55, nooijen93_15, nooijen95_1681,kowalski14_094102,kowalski16_144101,kowalski16_062512,kowalski18_4335,helgaker2014molecular, Schirmer2010BiCC} with respect to the $N$-electron single Slater determinant $|\phi_0^{(N)}\rangle$, the ground state $| \Psi^{(N)}\rangle$ and its adjoint $\langle \Psi^{(N)} |$ can be expressed as
\begin{align}
| \Psi^{(N)} \rangle &= \exp(T^{(N)})|\phi_0^{(N)}\rangle; \\
\langle \Psi^{(N)} | &= \langle \phi_0^{(N)} | (1+\Lambda^{(N)} ) \exp(-T^{(N)}),
\end{align}
the one-particle Green's function can be rewritten as
\textcolor{black}{
\begin{align}
    G_c^{\rm{R}}(t) =& -i \Theta(t) \exp(-iE_{CC}^{(N)} t)  \langle \phi_0^{(N)} | (1 + \Lambda^{(N)}) \times\notag \\
    & \exp(-T^{(N)}) a_c^\dagger \exp(iHt) \exp(T^{(N)})|\phi_0^{(N-1)}\rangle . \label{eq: chgf_cc} 
\end{align}
}
Here, the $N$-electron ground state energy $E^{(N)}_g$ is replaced by the corresponding CC energy $E^{(N)}_{CC}$. The excitation and de-excitation CC operators, $T^{(N)}$ and $\Lambda^{(N)}$, are defined as follows: 
\begin{align}
    T^{(N)} &= \sum_n t_n^{(N)} \mathbb{E}_n; \\
    \Lambda^{(N)} &= \sum_n l_n^{(N)} \mathbb{E}_n^\dagger
\end{align}
with $t_n$ and $l_n$ representing the amplitudes, and $\mathbb{E}_n$ and $\mathbb{E}_n^\dagger$ being the excitation and de-excitation generation operators, respectively. These operators are labeled by the compound index $n$ for single, double, or higher order excitations. For example, for single and double excitations,
\begin{align}
    &n^{(N)}=(p,q),~~|n^{(N)}\rangle = \mathbb{E}_n |\phi_0^{(N)}\rangle = a_p^\dagger a_q|\phi_0^{(N)}\rangle; \notag \\
    &n^{(N)}=(p,q,r,s),~~|n^{(N)}\rangle = \mathbb{E}_n |\phi_0^{(N)}\rangle = a_p^\dagger a_q^\dagger a_s a_r|\phi_0^{(N)}\rangle, \notag
\end{align}
here the indices $p,q,r,s\cdots$ label the spin-orbitals involved in the excitations. In Eq. (\ref{eq: chGF}), utilizing the commutative property $[a_c,T^{(N)}]=0$, it follows that
\begin{align}
    | \Psi_c^{(N-1)}\rangle = a_c \exp(T^{(N)})|\phi_0^{(N)}\rangle = \exp(T^{(N)})|\phi_0^{(N-1)}\rangle \label{eq: ans0}
\end{align}
with $|\phi_0^{(N-1)}\rangle = a_c |\phi_0^{(N)}\rangle$ being the $(N-1)$-electron single determinant. Throughout this paper, we use the superscript $(N)$ or $(N-1)$ in the operators and states for labeling the Hilbert space in which the operators and states are defined. 

\subsection{Time-dependent equation-of-motion coupled cluster ansatz}\label{sec: cc}

In our previous formulation the time-dependent equation-of-motion coupled cluster (EOM-CC) ansatz\cite{PhysRevB.18.6606,Rehr2020EOMCC} the $N$-electron state $| \Psi^{(N)}\rangle$, and therefore the $(N-1)$-electron state at $t=0$, $| \Psi_c^{(N-1)}(0)\rangle = | \Psi_c^{(N-1)}\rangle$, were assumed to be single Slater determinants. The time evolution of the $(N-1)$-electron state, $| \Psi_c^{(N-1)}(t)\rangle$, for any time $t$, is described by the following CC ansatz:
\begin{align}
    | \Psi_c^{(N-1)}(t)\rangle := N_c(t) \exp(T^{(N-1)}(t)) | \phi_0^{(N-1)}\rangle, \label{eq: ans1}
\end{align}
where $N_c(t)$ is a time-dependent normalization factor. The time-dependent CC operator $T^{(N-1)}(t)$, defined in the $(N-1)$-particle space with the initial condition $T^{(N-1)}(0)=0$, includes transitions from the occupied spin-orbitals to the hole, which are absent in $T^{(N)}$.

Integrating the CC ansatz in Eq. (\ref{eq: ans1}) into the time-dependent Schr\"{o}dinger equation (TDSE) and after some reformulations, we derive the equations-of-motion (EOMs) for the normalization factor $N_c(t)$ and the CC amplitudes $t^{(N-1)}_n(t)$:
\begin{align}
    \textcolor{black}{-}i\partial\ln N_c(t)/\partial t &= \langle \phi_0^{(N-1)} | \bar{H}(t) | \phi_0^{(N-1)} \rangle \notag \\
    &= E_{CC}^{(N-1)}(t), \\
    \textcolor{black}{-}i\partial t^{(N-1)}_n(t)/\partial t &= \langle n^{(N-1)} | \bar{H}(t) | \phi_0^{(N-1)} \rangle, \label{eq: ode_ans1}
\end{align}
where the time-dependent similarity transformed Hamiltonian $\bar{H}(t) = \exp(-T^{(N-1)}(t)) H \exp(T^{(N-1)}(t))$. The one-particle Green's function is then expressed as
\begin{align}
    G_c^{\rm{R}}(t) &= -i \Theta(t) \exp(-iE_{CC}^{(N)} t) N_c(t)  O(t) \notag \\
    &= -i \Theta(t) \exp(-iE_{CC}^{(N)} t) \exp(\int_0^t i E_{CC}^{(N-1)}(\tau)\rm{d}\tau)  \notag \\
    &= -i \Theta(t) \exp\big(-i \Delta E_{CC}(t) t \big) . \label{eq: chgf1}
\end{align}
where 
\begin{align}
N_c(t) &= N_c(0) \exp(\int_0^t i E_{CC}^{(N-1)}(\tau)\rm{d}\tau),
\end{align}
with $N_c(0) = 1$ and $\Delta E_{CC}(t) = E_{CC}^{(N)} - [E^{(N-1)}_{CC}]_t$. Here, $[E^{(N-1)}_{CC}]_t$ denotes the time average of $E^{(N-1)}_{CC}(\tau)$ over the period $t$, 
\begin{align}
[E^{(N-1)}_{CC}]_t = \frac{1}{t}\int_0^t E_{CC}^{(N-1)}(\tau)\rm{d}\tau. 
\end{align}
and
\begin{align}
O(t) &= \langle \phi_0^{(N)} | a_c^\dagger \exp(T^{(N-1)}(t)) | \phi_0^{(N-1)}\rangle = 1. \label{eq: ovlp0}
\end{align}
%

\subsection{Time-dependent double coupled cluster ansatz} \label{sec: dcc}

\begin{figure}
    \centering
    \includegraphics[width=\linewidth]{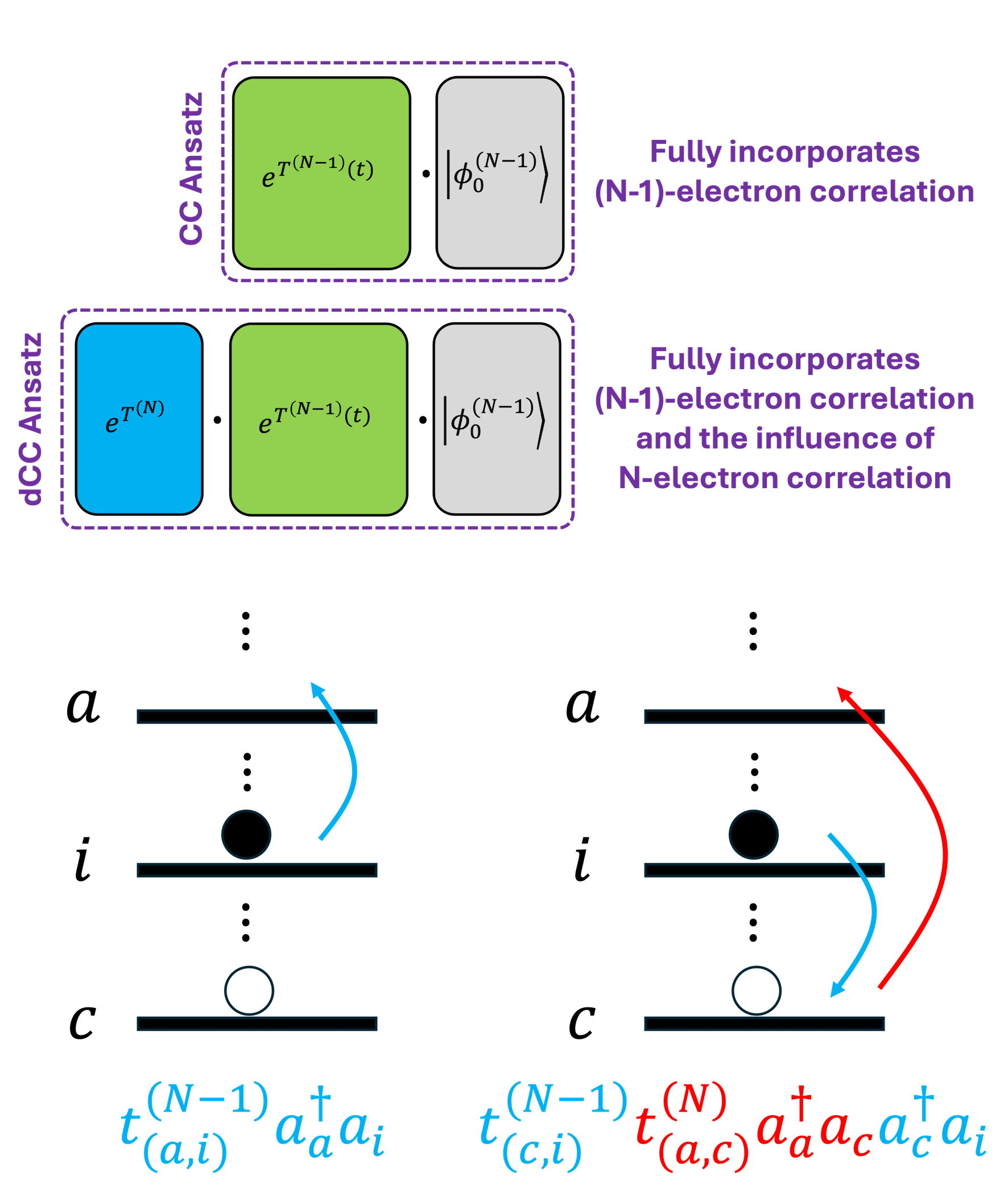}
    \caption{The comparison of ans\"{a}tze expressed in Eq. (\ref{eq: ans1}) and Eq. (\ref{eq: ans2}) in terms of correlation level (top) and the diagrams \textcolor{black}{(bottom)} representing the excitation from an occupied spin-orbital $i$ to a virtual spin-orbital $a$ following the generation of a hole at spin-orbital $c$ (bottom). The CC ansatz (\ref{eq: ans1}) is only able to describe the leading component of this excitation (bottom left), while the dCC ansatz (\ref{eq: ans2}) provides some higher order corrections, mediated through a hole at spin-orbital $c$ (bottom right).}
    \label{fig: ansatze comparison}
\end{figure}

It is worthwhile to notice that correlation effects in the $N$-electron state, $|\Psi^{(N)}\rangle$, as described in the exact one-particle Green's function formulation Eq. (\ref{eq: chGF}), was approximated by a single determinant as seen from the above. This distinction can also be explicitly observed by comparing the ans\"{a}tze in Eqs. (\ref{eq: ans0}) and (\ref{eq: ans1}). To enhance the description of correlation effects, particularly by incorporating correlations corresponding to the $N$-electron state, we propose a  more general time-dependent double CC (dCC) ansatz
\begin{align}
    | \Psi_c^{(N-1)}(t)\rangle := \tilde{N}_c(t) \exp(T^{(N)}) \exp(T^{(N-1)}(t)) | \phi_0^{(N-1)}\rangle \label{eq: ans2}
\end{align}
with $\tilde{N}_c(0) = 1$ and
\begin{align}
    | \Psi_c^{(N-1)}(0)\rangle = \exp(T^{(N)}) | \phi_0^{(N-1)}\rangle 
\end{align}
which reproduces ans\"{a}tze (\ref{eq: ans0}). The difference in correlation level between the two ans\"{a}tze (\ref{eq: ans1}) and (\ref{eq: ans2}) is summarized in the top of Figure \ref{fig: ansatze comparison}. Specifically, through the Taylor expansion of $\exp(T^{(N)})$, it is evident that the original CC ansatz (\ref{eq: ans1}) is the leading term of the dCC ansatz (\ref{eq: ans2}). Consequently, the dCC ansatz (\ref{eq: ans2}) is capable of capturing some hole-mediated higher-order corrections to the excitation following the generation of a hole at spin-orbital $c$, as illustrated at the bottom panel of Figure \ref{fig: ansatze comparison}. \textcolor{black}{We have also explored swapping the order of $e^{T^{(N-1)}}$ and $e^{T^{(N)}}$ in (\ref{eq: ans2}) to propose an alternative dCC ansatz. In Appendix \ref{sec: appA}, we compare both dCC ans\"{a}tze and demonstrate that the terms establishing the connection between $e^{T^{(N-1)}(t)}$ and $e^{T^{(N)}}$ arise exclusively in the dCC ansatz given by (\ref{eq: ans2}), making it the appropriate choice for exact Green's function calculations.}

\begin{figure}
    \centering
    \includegraphics[width=\linewidth]{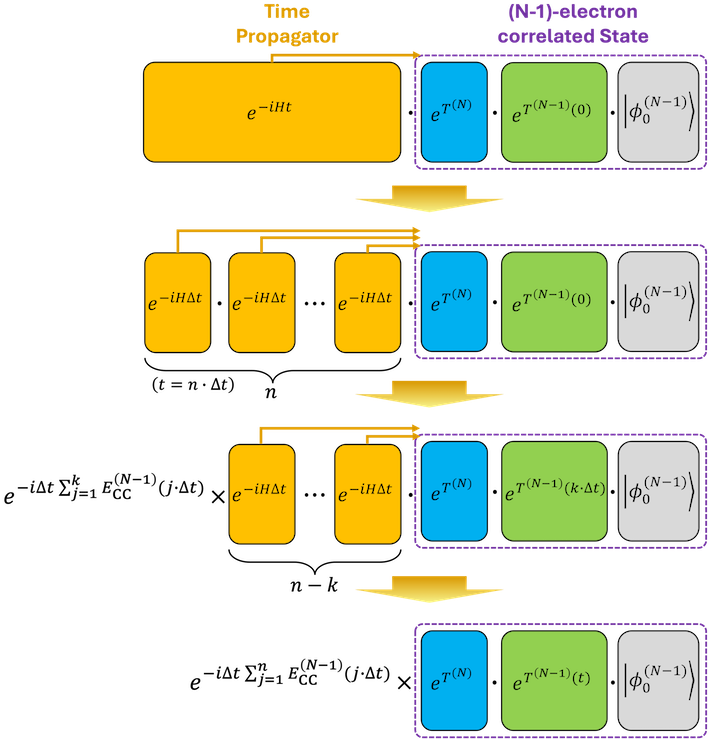}
    \caption{The schematic illustration of the new TDCC ansatz (\ref{eq: ans2}) and how it works to evaluate the time propagation of an $(N-1)$-electron correlated state.}
    \label{fig:workflow}
\end{figure}

The new dCC ansatz (\ref{eq: ans2}) bears resemblance to the double-unitary CC ansatz,\cite{Bauman2019SESDUCC} which incorporates two sets of excitation operators. When the dCC ansatz (\ref{eq: ans2}) is applied to the time-dependent Schr\"{o}dinger equation (TDSE), the EOMs for $\tilde{N}_c(t)$ and $t^{(N-1)}_n(t)$ are derived as follows:
\begin{align}
    \textcolor{black}{-} i\partial\ln \tilde{N}_c(t)/\partial t &= \langle \phi_0^{(N-1)} | \bar{\bar{H}}(t) | \phi_0^{(N-1)} \rangle \notag \\
    &= E_{dCC}^{(N-1)}(t), \label{eq: dCC1} \\
    \textcolor{black}{-} i\partial t^{(N-1)}_n(t)/\partial t &= \langle n^{(N-1)} | \bar{\bar{H}}(t) | \phi_0^{(N-1)} \rangle, \label{eq: dCC2}   
\end{align}
where the double similarity transformed Hamiltonian for the one-particle Green's function is defined as:
\begin{align}
 \bar{\bar{H}}(t) &= \exp(-T^{(N-1)}(t)) \exp(-T^{(N)}) H \cdot \notag \\
 &~~~~ \exp(T^{(N)}) \exp(T^{(N-1)}(t)). \label{eq: d_ST}
\end{align}
It is important to note that Eqs. (\ref{eq: dCC1}) and (\ref{eq: dCC2}) depend on solving the $N$-electron CC equations for $T^{(N)}$, followed by the utilization of the ordinary differential equation (ODE) integrator for the $(N-1)$-electron state to propagate $\tilde{N}_c(t)$ and $t^{(N-1)}_n(t)$. The workflow of the time propagation of the time-dependent $(N-1)$-electron correlated state described using the ansatz (\ref{eq: ans2}) is depicted in Figure \ref{fig:workflow}. Consequently, the one-particle Green's function $G_c^{\rm{R}}(t)$ can be reformulated as:
\begin{align}
G_c^{\rm{R}}(t) =& -i \Theta(t) \exp(-iE_{CC}^{(N)} t) \tilde{N}_c(t) \tilde{O}(t)  \label{eq: chgf2}
\end{align}
where 
\begin{align}
    \tilde{N}_c(t) &= \exp(i [E^{(N-1)}_{dCC}]_t t); \label{eq: N_c}
\end{align}
and the time-dependent overlap function $\tilde{O}(t)$ is formulated as
\begin{align}
\tilde{O}(t) &= \langle \phi_0^{(N)} | (1+\Lambda^{(N)}) \overline{a_c^\dagger} \exp(T^{(N-1)}(t)) | \phi_0^{(N-1)}\rangle. \label{eq: ovlp}
\end{align}
Note that, different from the previous approach, $\langle \Psi_c^{(N-1)}|$ is replaced by the correlated CC $\Lambda$-representation
\textcolor{black}{
\begin{align}
    \langle \Psi_c^{(N-1)} | = \langle \phi_0^{(N)} | (1+\Lambda^{(N)})  \overline{a_c^\dagger} \exp(-T^{(N)}),
\end{align}
}
from which the modified creation operator $\overline{a_c^\dagger}$ is given by 
\begin{align}
    \overline{a_c^\dagger} &= \exp(-T^{(N)}) a_c^\dagger \exp(T^{(N)})
    = a_c^\dagger + [a_c^\dagger, T^{(N)}].
\end{align}

It is worth noting that the dCC ansatz (\ref{eq: ans2}) involves the product of two exponential operators, leading to double similarity transformation (\ref{eq: d_ST}) in solving the EOMs (\ref{eq: dCC1}) and (\ref{eq: dCC2}). The double similarity transformation can potentially increase the non-linearity and leads to instability in the numerical propagation, thereby deteriorating the performance of the ODE integrator. In practical implementation, since the first similarity transformation, $\bar{H} = \exp(-T^{(N)})H\exp(T^{(N)})$ is time-independent, it can be computed upfront in the $N$-particle space before the time propagation in the $(N-1)$-particle space, with the computational cost paid for constructing $\bar{H}$. Alternatively, the construction of $\bar{H}$ can be entirely avoided by employing approximate ans\"{a}tze that combine the product of two exponential operators into one.

Straightforward approximations can be obtained by utilizing the Baker–Campbell–Hausdorff formula in (\ref{eq: ans2}) and truncating the expansion at different (commutator) levels, for example:
\begin{align}
    | \Psi_c^{(N-1)}(t)\rangle &\approx \tilde{N}_c(t) \times \exp(T^{(N)}+T^{(N-1)}(t)) | \phi_0^{(N-1)}\rangle, \label{eq: ans3} \\
    | \Psi_c^{(N-1)}(t)\rangle &\approx \tilde{N}_c(t) \times \exp\left(T^{(N)}+T^{(N-1)}(t) \right. \notag \\
     &~~ \left. +\frac{1}{2}[T^{(N)},T^{(N-1)}(t)] \right) | \phi_0^{(N-1)}\rangle, \label{eq: ans4} \\
     & \vdots \notag
\end{align}
We denote the approximate dCC ans\"{a}tze (\ref{eq: ans3},\ref{eq: ans4}) as the dCC-1 and dCC-2 ans\"{a}tze, respectively. The benefit of using the approximate dCC ans\"{a}tze in the EOMs (\ref{eq: dCC1} and \ref{eq: dCC2}) is to re-utilize the conventional CC implementation with modest modifications in the lists of the CC excitation operators. We will examine the numerical performance of such approximations in model systems in the following section.


\begin{figure*}[!htbp]
    \centering
    \includegraphics[width=\linewidth]{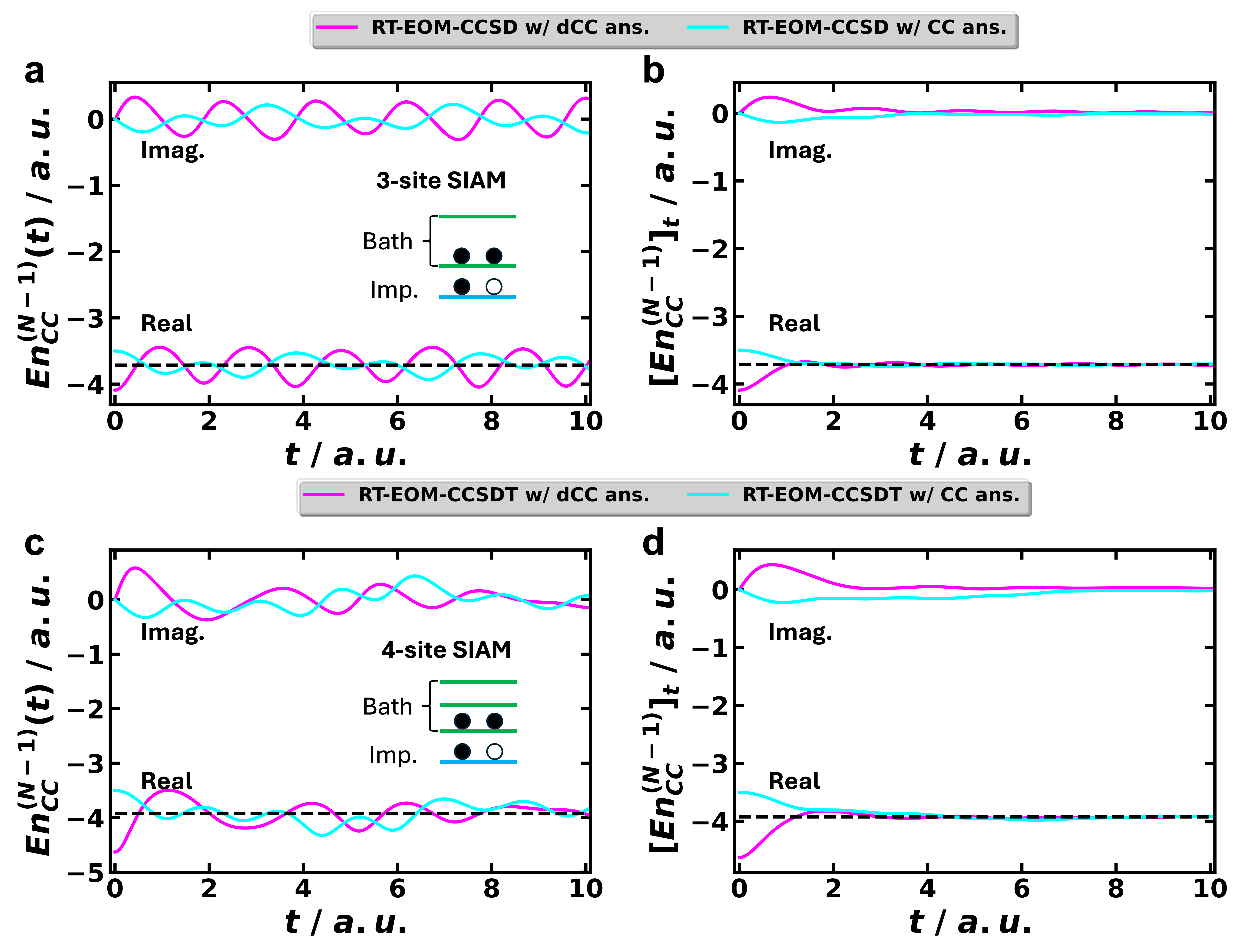}
    \caption{\textcolor{black}{(a,c)} The energy fluctuation of the $(N-1)$-electron states of the three-site \textcolor{black}{(a)} and four-site  \textcolor{black}{(c)} SIAMs with $U=3.0$ a.u. in the RT-EOM-CC simulations within $t\in[0,10]$ a.u. employing the ans\"{a}tze expressed in Eq. (\ref{eq: ans1}) and Eq. (\ref{eq: ans2}). The reference state in the RT-EOM-CC simulations is a three-electron state (see the insets \textcolor{black}{(a,c)}) where one electron resides at impurity site and two at the bath sites. \textcolor{black}{In (b), the time-average of the time-dependent $(N-1)$-electron energy, $[E_{CC}^{(N-1)}]_t$, of the three-site SIAM converges to the stationary $(N-1)$-electron CCSD energies (denoted by the dashed lines), $E_{CCSD}^{(N-1)}$ of $-3.713370$ a.u. In (d), $[E_{CC}^{(N-1)}]_t$ of the four-site SIAM converges to $E_{CCSDT}^{(N-1)}$ of $-3.925962$ a.u. In (b,d), a slight difference in the convergence performance of the RT-EOM-CCSD approaches using different ansätze is observed; a preliminary discussion is provided in Appendix \ref{sec: appB}.}}
    \label{fig: En_SIAM}
\end{figure*}

\section{Numerical results and discussion}

In this paper, we focus on evaluating the performance of our newly proposed ansatz for computing the one-particle Green's function in its exact limit. Particularly, we compare it with the performance of the previous ansatz. To this end, we employ the SIAM as our test framework within the RT-EOM-CC approach to compute the exact one-particle impurity Green's function. The SIAM Hamiltonian is expressed as follows:
\begin{align}
H_{\rm SIAM} = H_{\rm imp.} + H_{\rm bath} + H_{hyb.}, \label{Hsiam}
\end{align}
where 
\begin{align}
H_{\rm imp.} = \sum_{\sigma} \mu_c c^\dagger_{\sigma} c_{\sigma} + U c^\dagger_{\uparrow}c_{\uparrow}c^\dagger_{\downarrow}c_{\downarrow} \label{Himp}
\end{align}
describes the impurity-site with potential $\mu_c$ and the Coulomb interaction $U$ between electrons with opposite spins ($\sigma = \uparrow$ or $\downarrow$) at the impurity site,
\begin{align}
H_{\rm bath} = \sum_{i=1,\sigma}^{N_{\rm b}} \mu_{d,i} d^\dagger_{i,\sigma} d_{i,\sigma} \label{Hbath}
\end{align}
characterizes the non-interacting bath sites with potentials $\mu_{d}$'s, and
\begin{align}
H_{\rm hyb.} = \sum_{i=1,\sigma}^{N_{\rm b}} V_i \big( c^\dagger_{\sigma} d_{i,\sigma} + d^\dagger_{i,\sigma} c_{\sigma} \big) \label{Hhyb}
\end{align}
describes the coupling between the impurity site and the bath sites due to the hybridization. 

In subsequent tests, we focus on the three-site and four-site SIAM configurations, setting $N_{b}=4$, $\mu_c = -1.5$ a.u., $V_i = 0.5$ a.u. $\forall i$, and $\mu_{d,i} \in [-1.0, 1.0]$ a.u. For the three-site SIAM, we employ the RT-EOM-CCSD approach with the two TDCC ans\"{a}tze described in the previous section to compute the one-particle Green's function under three Coulomb interactions, $U \in \{1.0, 2.0, 3.0\}$ a.u. The purpose is to study how the RT-EOM-CCSD approach with different ans\"{a}tze behaves under different on-site correlation levels and compare it to the exact solution. The exact one-particle Green's function were obtained by Eq. (\ref{eq: chGF}) employing the exact diagonalization of the Hamiltonian. For the four-site SIAM, we employ the RT-EOM-CCSDT approach with the two TDCC ans\"{a}tze to compute the one-particle Green's function with the on-site Columb interaction $U=3.0$ a.u. to study the performance difference between the two TDCC ans\"{a}zte in the RT-EOM-CC simulation with increased theoretical level.

In all the RT-EOM-CC simulations, the $N$-electron CC operators were obtained from converged CC ground state calculations with the convergence criteria of the energy change being less than $10^{-6}$ a.u. and the norm of the CC amplitude change being less than $10^{-7}$. The Runge-Kutta-Fehlberg approach, RK45, and its implementation in SciPy\cite{2020SciPy-NMeth} were used to numerically solve the ODEs (\ref{eq: ode_ans1}) and (\ref{eq: dCC2}) for obtaining $G_c^R(t)$ with $t \in [0,250]$ a.u., unless otherwise mentioned.


\subsection{Energy fluctuations of $(N-1)$-electron states}

We first evaluate the energy fluctuation of the $(N-1)$-electron state in the RT-EOM-CC simulations. For a three-site SIAM, as depicted in the inset of Figure \ref{fig: En_SIAM}a, since the highest level of excitations are double excitations, the exact propagation of the non-equilibrium $(N-1)$-electron state $|\Psi_c^{(N-1)}\rangle$ can be accurately performed at the coupled cluster with singles and doubles (CCSD) level. Figure \ref{fig: En_SIAM}a illustrates the energy fluctuations of the $(N-1)$-electron state in RT-EOM-CCSD simulations using both the previous ansatz (\ref{eq: ans1}) and the new ansatz (\ref{eq: ans2}). In both ans\"{a}tze, the CCSD operators are defined as:
\begin{align}
T^{(N/N-1)} &= T_1^{(N/N-1)} + T_2^{(N/N-1)}, \notag \\
\Lambda^{(N)} &= \Lambda_1^{(N)} + \Lambda_2^{(N)}.
\end{align}
Notably, the two RT-EOM-CCSD simulations have different energy starting points due to the choice of the ansatz.  The starting energy in the simulation using ansatz (\ref{eq: ans1}) is the Hartree-Fock energy of the $(N-1$-electron state, while the starting energy with ansatz (\ref{eq: ans2}) is considered to be that of a non-stationary $(N-1)$-electron correlated state. Additionally, as shown Figure \ref{fig: En_SIAM}b, although the time-dependent energy curves of the $(N-1)$-electron state vary between the two simulations, the time-averaged energies eventually converge to the same stationary $(N-1)$-electron CCSD energy of the half-filled three-site SIAM, irrespective of the ansatz used. 

Figure \ref{fig: En_SIAM}c,d display the energy fluctuations in the RT-EOM-CCSDT simulations of the four-site SIAM. Due to the additional bath site compared to the three-site model, the CC operators in the exact limit are expanded as follows:
\begin{align}
T^{(N)} &= T_1^{(N)} + T_2^{(N)} + T_3^{(N)} + T_4^{(N)}, \notag \\
T^{(N-1)} &= T_1^{(N-1)} + T_2^{(N-1)} + T_3^{(N-1)}, \notag \\
\Lambda^{(N)} &= \Lambda_1^{(N)} + \Lambda_2^{(N)} + \Lambda_3^{(N)} + \Lambda_4^{(N)}.
\end{align}
Although the energy fluctuations are not as uniform as in the three-site model, the similar converging behavior of the time-averaged energies is observed, albeit over a slightly longer duration. It is worth noting that, with the selected parameters, the impact of quadruple excitations on the exact ground state of the four-site SIAM becomes negligible, therefore the exact ground state  can be well approximated by the CCSDT wave function with an energy deviation $<2.0 \times 10^{-7}$ a.u. 

\subsection{$G_c^R$ computed by different RT-EOM-CC approaches}

\begin{figure}
    \centering
    \includegraphics[width=\linewidth]{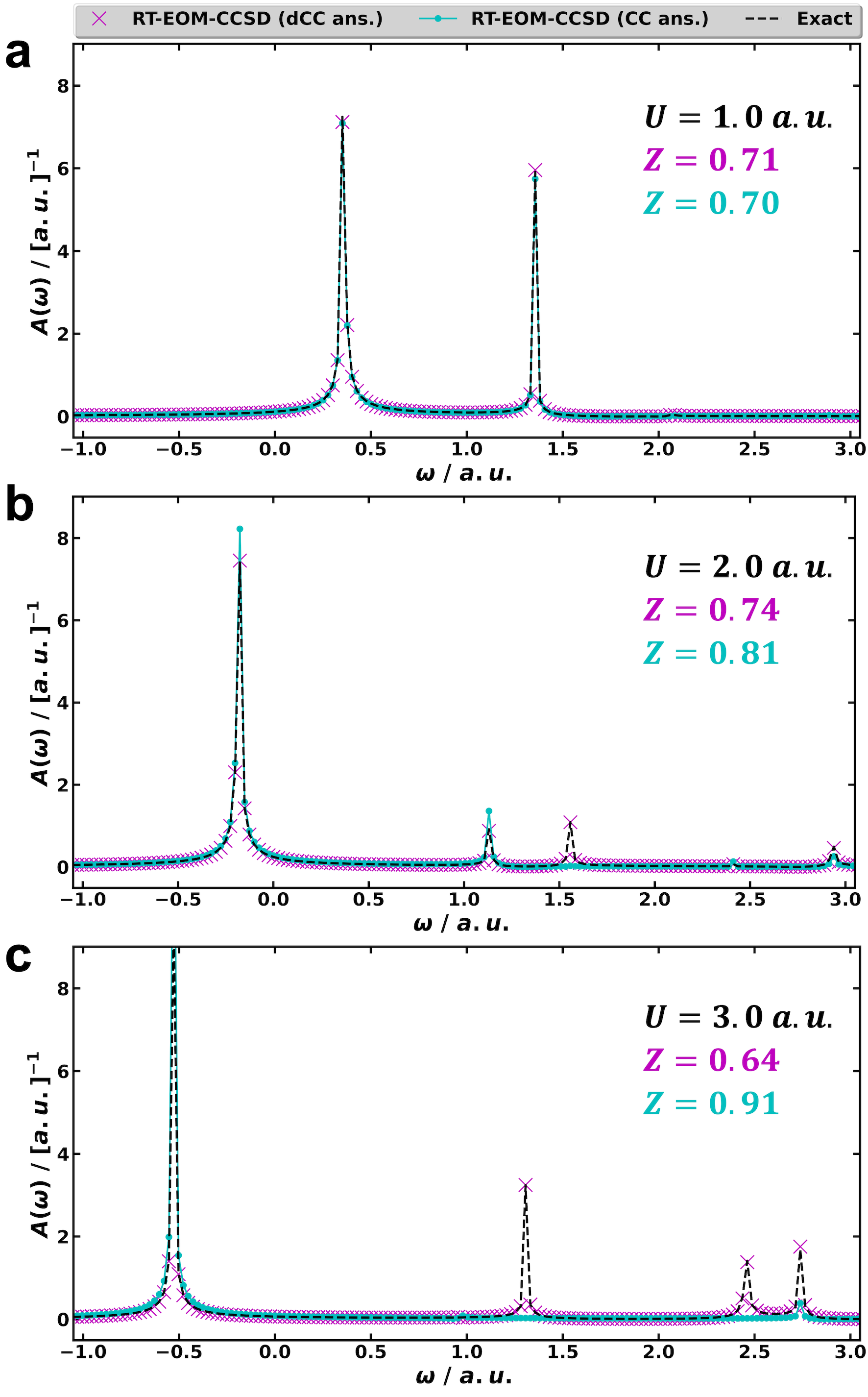}
    \caption{The spectral functions, $A(\omega)$, of the three-site SIAM ($N=4$) with different values of the Coulomb interaction $U$, \textcolor{black}{(a) $U=1.0$ a.u., (b) $U=2.0$ a.u., and (c) $U=3.0$ a.u.,} computed using the RT-EOM-CCSD approach employing the two ans\"{a}tze in Eqs. (\ref{eq: ans1}) and (\ref{eq: ans2}). In Eq. (\ref{eq: ans2}), the $N$-electron CC amplitude, $T^{(N)}$, is obtained from the $N$-electron CCSD ground state calculation of the three-site SIAM. The exact curve is obtained through the exact diagonalization of the Hamiltonian. The strength of the main peak in each computed spectral function is given by the renormalization constant $Z$ (see Appendix \ref{app:G_cumulant} for the computation details). The broadening factors in the computed spectral functions are $\eta=0.016$ a.u. ($U=1.0$ a.u.), $\eta=0.016$ a.u. ($U=2.0$ a.u.), and $\eta=0.0095$ a.u. ($U=3.0$ a.u.), respectively.}
    \label{fig:Gfxn_3site}
\end{figure}

\begin{figure}
    \centering
    \includegraphics[width=\linewidth]{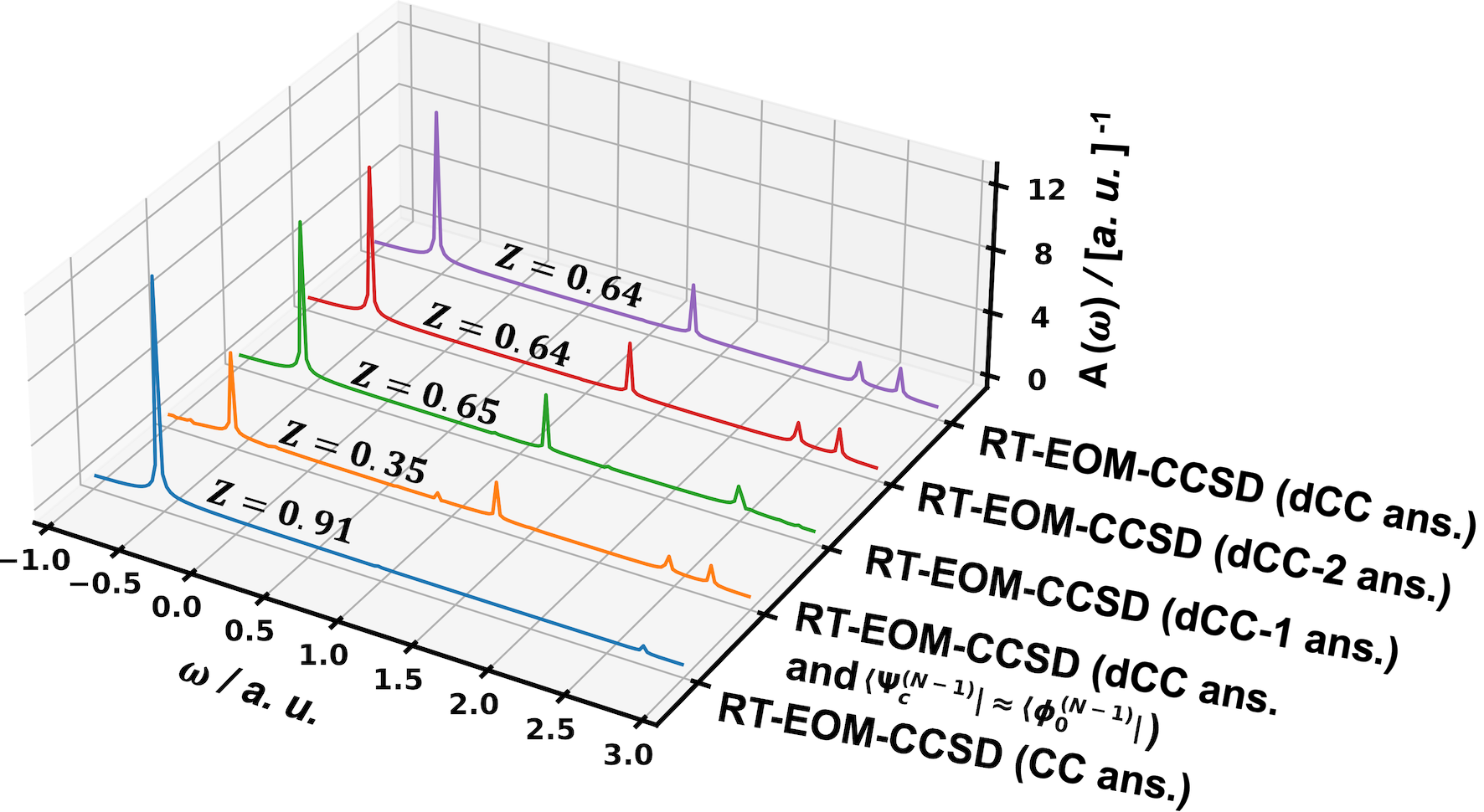}
    \caption{The spectral function, $A(\omega)$, of the three-site SIAM ($N=4$) with the Coulomb interaction $U=3.0$ a.u. computed using the RT-EOM-CCSD approach employing the different ans\"{a}tze and approximations. The strength of the main peak in each computed spectral function is given by the renormalization constant $Z$. The broadening factor in all the computed spectral functions is $\eta=0.0095$ a.u.}
    \label{fig:Gfxn_3site_diff_ans}
\end{figure}

\begin{figure}
    \centering
    \includegraphics[width=\linewidth]{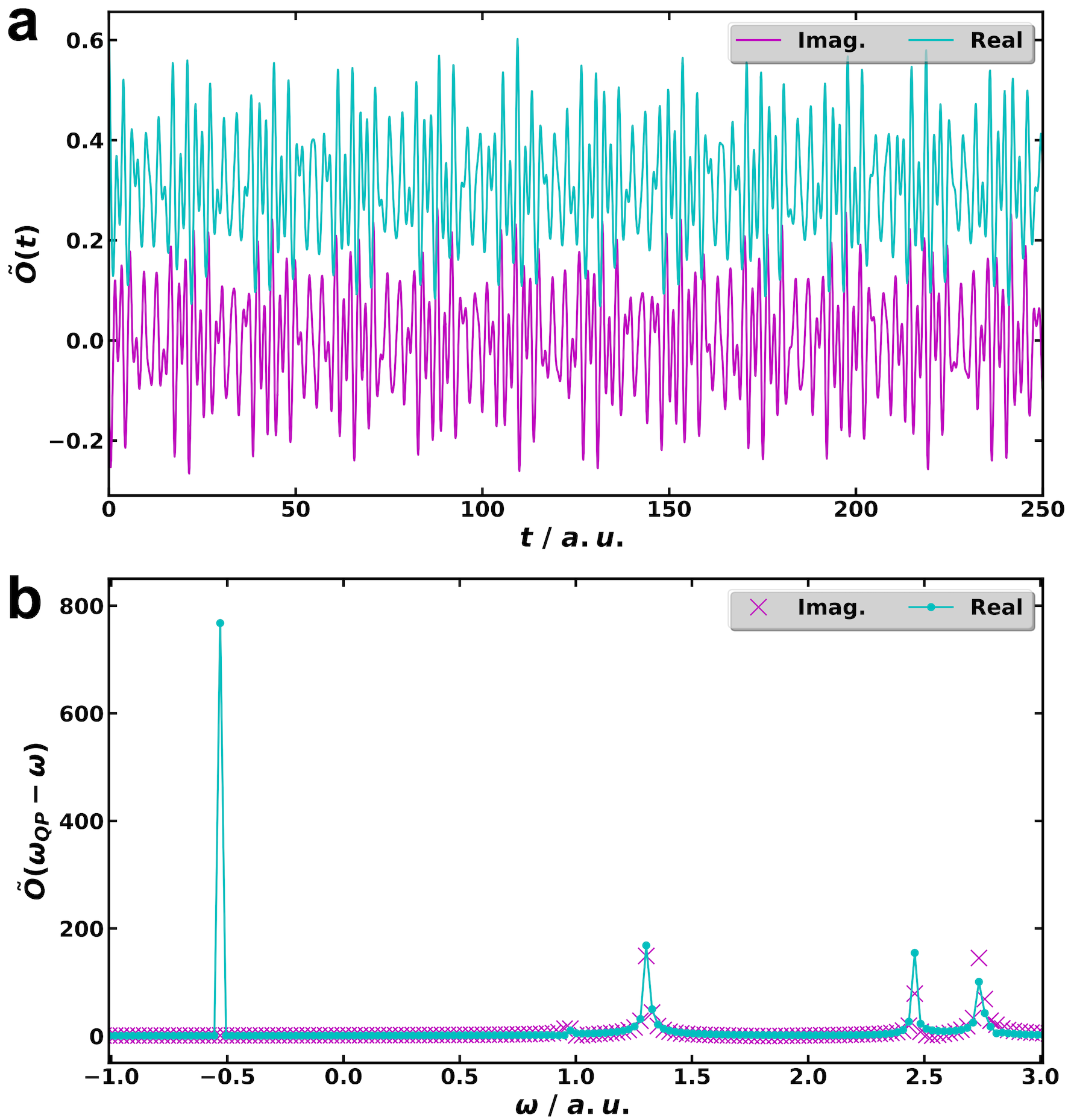}
    \caption{The overlap function in the \textcolor{black}{(a) time and (b) frequency} domains employed in the RT-EOM-CCSD Green's function simulation of the three-site SIAM.}
    \label{fig:Gfxn_3site_overlap_effect}
\end{figure}

\begin{figure}
    \centering
    \includegraphics[width=\linewidth]{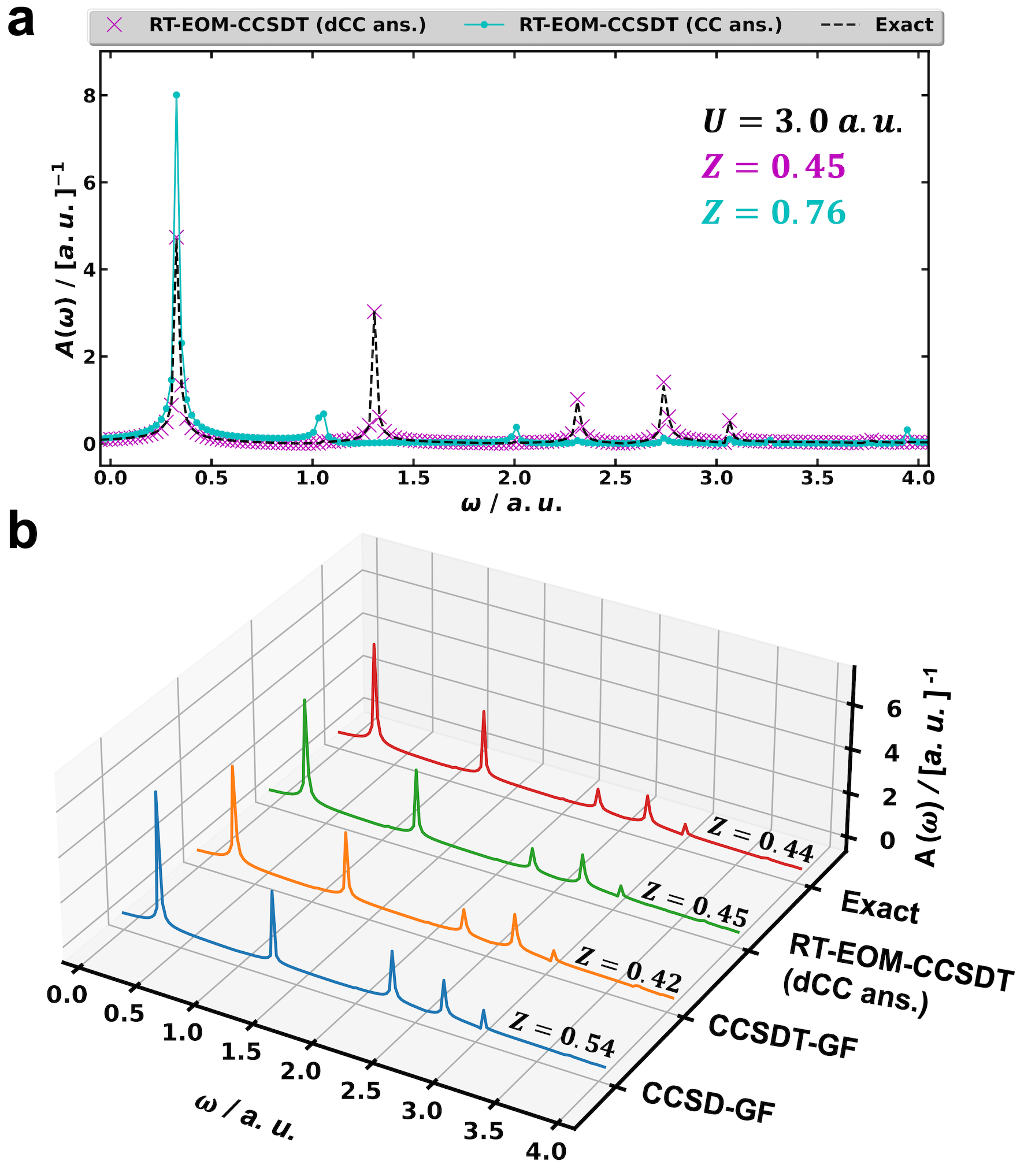}
    \caption{\textcolor{black}{(a)} The spectral functions, $A(\omega)$, of the four-site SIAM ($N=4$, $U=3.0$ a.u.) computed using the RT-EOM-CCSDT approach employing the two ans\"{a}tze (\ref{eq: ans1}) and (\ref{eq: ans2}). In the ansatz (\ref{eq: ans2}), the $N$-electron CC amplitude, $T^{(N)}$, is obtained from the $N$-electron CCSDTQ calculation of the four-site SIAM. (b) The comparison of the spectral functions of the four-site SIAM ($N=4$) computed by different theoretical approaches. The exact values are obtained from the exact diagonalization of the Hamiltonian, and can also be equivalently obtained through CCSDTQ-GF method. The strength of the main peak in each computed spectral function is given by the renormalization constant $Z$. The broadening factor in all the computed spectral functions is $\eta=0.015$ a.u.}
    \label{fig:Gfxn_4site_ccsdt}
\end{figure}

We then proceed to examine the computed $G_c^R$ from the different RT-EOM-CC approaches presented in the previous section.
Figure \ref{fig:Gfxn_3site} presents the impurity Green's function of the three-site SIAM, computed using the previous and present RT-EOM-CCSD approaches, where the previous approach employs the CC ansatz (\ref{eq: ans1}) and the approximation $\langle \Psi_c^{(N-1)} | \approx \langle \phi_0^{(N-1)}|$, while the present approach employs the dCC ansatz (\ref{eq: ans2}) and represents $\langle \Psi_c^{(N-1)}$ using the $\Lambda$-CC formulation. For comparison, the exact $G_{c}^{\rm R}$ curves, computed by Eq. (\ref{eq: chGF}) and through exact diagonalization of the Hamiltonian, are also provided. As shown in Figure \ref{fig:Gfxn_3site}, the present RT-EOM-CCSD approach that employs the dCC ansatz (\ref{eq: ans2}) successfully reproduces the exact $G_c^R(t)$ curves regardless of the strength of the Coulomb interaction. On the other hand, the previous RT-EOM-CCSD approach accurately reproduces the exact curve only when the Coulomb interaction is relatively small (e.g. $U=1.0$ a.u., as seen in Figure \ref{fig:Gfxn_3site}a). However, when the Coulomb interaction is stronger, as illustrated in Figure \ref{fig:Gfxn_3site}b,c, the spectral function computed by the previous RT-EOM-CCSD approach captures only the main (quasi-particle) peak and misses some satellite peaks. The performance difference between two RT-EOM-CC approaches is also reflected in the numerical difference in the renormalization constant $Z$ of the computed spectral functions. The $Z$ value, which lies between zero and one, is often used to quantify the strength of the main peak in the computed spectral function, with $Z\rightarrow 0$ indicating stronger many-body interactions that lead to more significant satellite features, and $Z\rightarrow 1$ indicating weaker interactions. As shown by the $Z$ values in Figure \ref{fig:Gfxn_3site}, the discrepancy in $Z$ between the two RT-EOM-CCSD approaches increases as the Coulumbic interaction strength increases, from $\Delta Z = \| Z_{dCC} - Z_{CC}\| \approx 0.01$ when $U=1.0$ a.u., to $\Delta Z \approx 0.07$ when $U=2.0$ a.u. and $\Delta Z \approx 0.27$ when $U=3.0$ a.u., underscoring the necessity of employing the present RT-EOM-CC approach in more strongly correlated cases. 

The detailed comparison of the computed spectral functions using various RT-EOM-CCSD approaches, employing different ans\"{a}tze and approximations, is presented in Figure \ref{fig:Gfxn_3site_diff_ans} for the three-site SIAM. Notably, two differences exist between the RT-EOM-CC approaches described in Sections \ref{sec: cc} and \ref{sec: dcc}: the TDCC ans\"{a}tze and the inclusion of the overlap function. The latter difference hinges on whether the approximation of the left eigenvector, $\langle \Psi_c^{(N-1)} | \approx \langle \phi_0^{(N-1)}|$, is applied. Notably, with this approximation, the renormalization constant $Z$, is either too high ($Z=0.91$ using the CC ansatz) or too low ($Z=0.35$ using the dCC ansatz) compared to the exact value ($Z=0.64$).  To closely examine the effect of these two factors, three additional approximations are also included in Figure \ref{fig:Gfxn_3site_diff_ans}. It is observed that for the RT-EOM-CCSD approach with the dCC ansatz, the computed spectral functions (orange and purple curves) are relatively insensitive to the inclusion of the approximation, $\langle \Psi_c^{(N-1)} | \approx \langle \phi_0^{(N-1)}|$. Specifically, the differences between two computed spectral functions lie only lies in the slightly varied intensities of the quasi-particle peaks and an insignificant artificial satellite between $[0.5,1.0]$ a.u. On the other hand, the spectral functions obtained with the RT-EOM-CCSd approaches using approximated dCC ans\"{a}tze and the explicit $\Lambda$-CC formulation of $\langle \Psi_c^{(N-1)} |$ exhibit the dependence on the truncation level in the BCH expansion. As shown by the green and red curves in Figure \ref{fig:Gfxn_3site_diff_ans}, employing the dCC-1 ansatz results in the omission of one satellite between 2.5 and 3.0 a.u. missing in the computed spectral function, whereas including the single commutator in the truncated BCH expansion, i.e. the dCC-2 ansatz, reproduces this missing peak and improves the overall agreement with respect to the exact spectral functions.

To evaluate the impact of overlap function on the computed Green's function, we plot $\tilde{O}(t)$ and its Fourier transform $\tilde{O}(\omega)$ in Figure \ref{fig:Gfxn_3site_overlap_effect}. As shown, the real part of $\tilde{O}(t)$ oscillates around 0.3, while the imaginary part oscillates around 0.0 over the simulation time (up to 250 a.u.). Therefore, neglecting $\tilde{O}(t)$ in the Green's function calculation, essentially assuming $\langle \Psi_c^{(N-1)} | = \langle \phi_0^{(N-1)}|$, results in a rough approximation (especially in more correlated scenarios). After incorporating Eq. (\ref{eq: ovlp}) into Eq. (\ref{eq: chgf2}), the Fourier transform of $G_c^{\rm R}(t)$ takes a convoluted form, where both the real and imaginary parts of the Fourier transform of $\tilde{O}(t)$ begin to influence the spectral function. Figure \ref{fig:Gfxn_3site_overlap_effect}b displays the shifted Fourier transform of $\tilde{O}(t)$, where the frequency shift corresponds to the position of the quasi-particle peak, $\omega_{QP}\approx -0.5308$ a.u., in the computed spectral function (see Figure \ref{fig:Gfxn_3site_diff_ans}). This shifted Fourier transform, $\tilde{O}(\omega_{QP}-\omega)$, aligns the peak positions with those in the computed spectral function (e.g. the purple peaks in Figure \ref{fig:Gfxn_3site_diff_ans}).

The performance difference between the two versions of the RT-EOM-CC approaches is also demonstrated in Figure \ref{fig:Gfxn_4site_ccsdt} for computing $G_c^R(\omega)$ of the four-site SIAM, incorporating up to triple excitations within the RT-EOM-CC framework. As shown in Figure \ref{fig:Gfxn_4site_ccsdt}a, the spectral function computed by the RT-EOM-CCSDT approach using the dCC ansatz (\ref{eq: ans2}) and the $\Lambda$-CC formulation of $\langle \Psi_c^{(N-1)} |$ reproduces the exact solution. On the other hand, the RT-EOM-CCSDT approach that employs the CC ansatz (\ref{eq: ans1}) combined with the approximation $\langle \Psi_c^{(N-1)} | \approx \langle \phi_0^{(N-1)}|$ locates only the main peak below 0.5 a.u., with predicted satellites either missing or red-shifted relative to the exact solution. Figure \ref{fig:Gfxn_4site_ccsdt}b compares the spectral function computed using the current RT-EOM-CCSDT approach with the dCC ansatz against those computed by the CC-GF approaches, including the CCSD-GF and CCSDT-GF methods. Here, the CC-GF values are obtained by substituting the CC left and right wave functions, $\langle \phi_0^{(N)}| (1+\Lambda^{(N)})\exp(-T^{(N)})$ and $\exp(T^{(N)})|\phi_0^{(N)}\rangle$, for $\langle \Psi^{(N)}|$ and $|\Psi^{(N)}\rangle$ in Eq. (\ref{eq: chGF}), respectively. For all the spectral functions computed by the RT-EOM-CC and CC-GF approaches, the ground state is obtained through the CCSDTQ calculation. The spectral functions computed by all three methods closely align with the exact solution, and the renormalization constants, $Z$, of the main peak are consistent with the exact value of 0.44. In particular, the CCSDT-GF and RT-EOM-CCSDT with the dCC ansatz (\ref{eq: ans2}), both offering essentially exact treatments of the four-site SIAM, accurately reproduce the exact solution in terms of peak positions and amplitudes. 

\subsection{Component analysis of $G_c^R$}

\begin{figure}
    \centering
    \includegraphics[width=\linewidth]{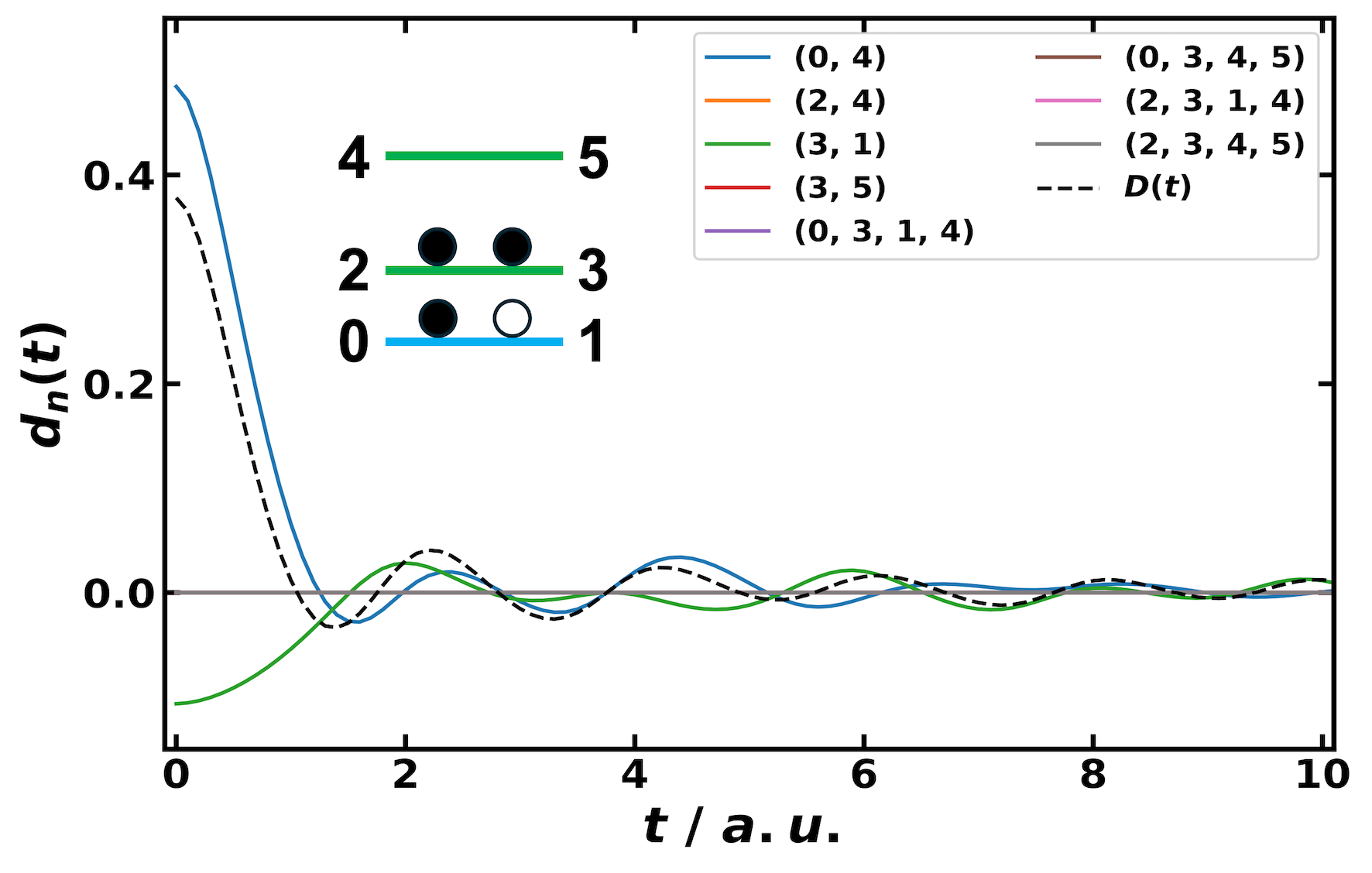}
    \caption{The time evolution of $D(t)$ and its leading components, and their Fourier transform, for computing the impurity Green's functions, $G_c^R(t)$ and $G_c^R(\omega)$, of the three-site SIAM ($U=3.0$ a.u.). The component analysis of $D(t)$ is based on the cluster analysis in Eqs. (\ref{eq: D_t}-\ref{eq: d_n_t}). The inset exhibits the schematic electron occupation of the three-site SIAM with one electron removed from the impurity site. The curves are labeled through the tuple $(i, a)$ or $(i,j,a,b)$ implying the energy contribution to $D(t)$ from the single excitation from the occupied spin-orbital $i$ to the unoccupied spin-orbital $a$, or from the double excitation from the occupied spin-orbitals $i,j$ to the unoccupied spin-orbitals $a,b$. }
    \label{fig:Gfxn_D}
\end{figure}

\begin{figure}
    \centering
    \includegraphics[width=0.9\linewidth]{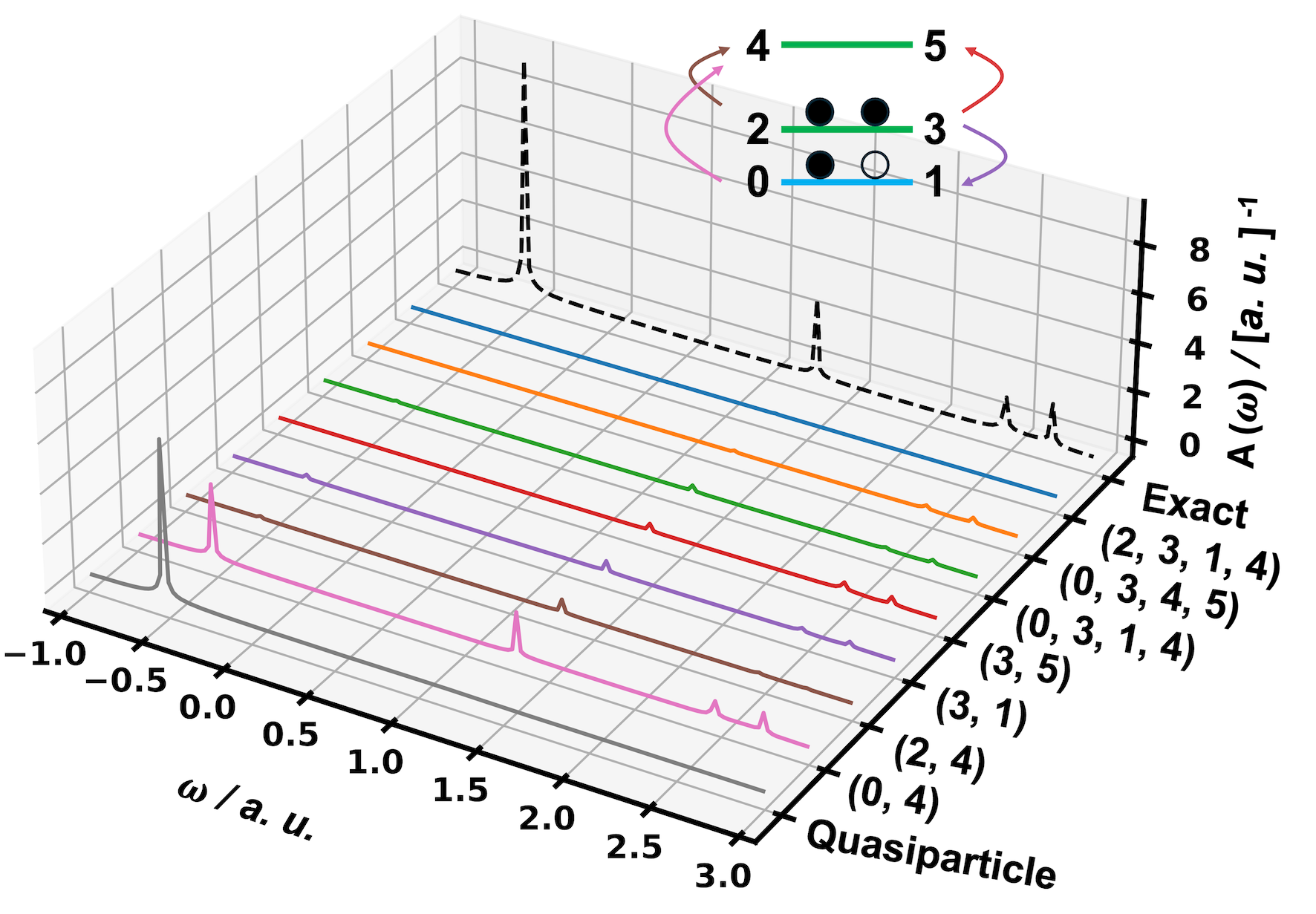}
    \caption{The approximate component analysis of the impurity Green's functions, $G_c^R(\omega)$, of the three-site SIAM ($U=3.0$ a.u.), employing the cluster analysis of $\tilde{\mathcal{O}}^*(\omega)$ in Eqs. (\ref{eq: ovlp_fft}-\ref{eq: T_fft}). }
    \label{fig:Gfxn_comp}
\end{figure}

Finally, we demonstrate the component analysis on the $G_c^R$ of the three-site SIAM employing the approach described in Appendices~\ref{sec: decomp} and \ref{sec: G_workingEq}. This analysis involves the decomposition of two terms, $\tilde{O}(t)$ and $A(t)$, and the convolution of their Fourier transform in the frequency domain contributes to both the main peaks and satellites of $G^R_c(\omega)$. Specially, for $A(t)$, according to Eq. (\ref{eq: FT_W}), the decomposition is fundamentally an analysis of $\exp(-iD(t)t)$, or $D(t)$. Figure \ref{fig:Gfxn_D} illustrates the time evolution of $D(t)$ and its components as detailed in Eqs. (\ref{eq: D_t}) and (\ref{eq: d_n_t}). The shifted Fourier transforms of the time evolutions of these components reproduce the RT-EOM-CCSD spectral function using ansatz (\ref{eq: ans2}), as indicated by the orange curve in Figure \ref{fig:Gfxn_3site_diff_ans}. Notably, only two excitations$-0\rightarrow 4$ and $3\rightarrow 1-$along with the removal of the electron at the spin-orbital \#$1$, contribute to both the main and satellite peaks in a convoluted manner, attributed to the exponential operation on $d_n(t)$'s. 
Regarding $\tilde{O}(t)$, the analysis is directly conducted through the cluster analysis described in Eqs. (\ref{eq: eT_expansion}-\ref{eq: T_fft}). It is worth mentioning that in the long time limit, if $A(t)$ is approximated only by the main peak (i.e., only one term in the expansion (\ref{eq: W_fitting})), then the component analysis of $G_c^R(t)$ can be approximately performed through the component analysis of $\tilde{O}(t)$, with results displayed in Figure \ref{fig:Gfxn_comp}. As shown, besides the quasi-particle peak at $\sim-0.53$ a.u., the leading excitations contributing to the main and satellite peaks include single excitations such as $0\rightarrow 4$ (across all peak positions), $2\rightarrow 4$, $3\rightarrow 1$ and $3\rightarrow 5$ (mainly at satellite positions between 1 and 3 a.u.), as well as the double excitation $0,3\rightarrow 1,4$ (at satellite positions between 1 and 3 a.u.).

\section{Conclusion and outlook}
In the paper, we have analyzed and examined a series of TDCC ans\"{a}tze in the RT-EOM-CC simulations for computing the one-particle Green's function. Unlike the previous CC ansatz used in RT-EOM-CC simulations, we introduced a new ansatz that features a double CC form$-$the product of the exponential CC operators from $N$ and $(N-1)$-particle spaces. Preliminary analysis and simulations on simple SIAMs demonstrate that, compared to the previous ansatz, the new dCC ansatz is capable of approaching the exact limit by incorporating hole-mediated higher order excitations in the $(N-1)$-electron CC exponential operator and by using small time steps. By employing the BCH expansion of the new dCC ansatz and truncating at different commutator levels, we have also introduced some approximate TDCC ans\"{a}tze to the RT-EOM-CC simulations. The approximate ans\"{a}tze feature a single exponential algebraic structure that potentially balances the complexity of implementation with accuracy. Additionally, we have formalized a recipe for analyzing the components of the computed Green's function in RT-EOM-CC simulations, paving the way for larger-scale and efficient implementations and detailed spectral function analysis for complex molecular systems in the near future. Future work will focus on incorporating the double-unitary CC ans\"{a}tze\cite{Bauman2019SESDUCC} into the RT-EOM-CC framework, extending RT-EOM-CC to compute the nonequilibrium Green's function,\cite{doi:10.1021/acs.jctc.8b00773,doi:10.1021/acs.jctc.9b00750}and improving numerical aspects including more stable ODE integrator\cite{doi:10.1021/acs.jctc.3c00911} and robust interpolation and extrapolation techniques\cite{peng2019mor,PhysRevB.107.075107}.

\section{Acknowledgements}

This material is based upon work supported by the ``Transferring exascale computational chemistry to cloud computing environment and emerging hardware technologies (TEC$^4$)''  project, which is funded by the U.S. Department of Energy, Office of Science, Office of Basic Energy Sciences, the Division of Chemical Sciences, Geosciences, and Biosciences (under FWP 82037). 
F.D.V and J.J.R. acknowledge the support from the Center for Scalable Predictive methods for Excitations and Correlated phenomena (SPEC), which is funded by the U.S. Department of Energy (DoE), Office of Science, Office of Basic Energy Sciences, Division of Chemical Sciences, Geosciences and Biosciences as part of the Computational Chemical Sciences (CCS) program at Pacific Northwest National Laboratory (PNNL) under FWP 70942. 
B.P. also acknowledges Dr. Niri Govind for the fruitful discussion during the preparation of the manuscript.


\appendix

\section{\textcolor{black}{Comparing $e^{T^{(N)}}e^{T^{(N-1)}}|\phi_0^{(N-1)}\rangle$ with $e^{T^{(N-1)}}e^{T^{(N)}}|\phi_0^{(N-1)}\rangle$}}\label{sec: appA}

\textcolor{black}{When proposing the dCC ansatz (\ref{eq: ans2}) for calculating the exact Green's function, a notable concern is the ordering of the two exponential operators. Intriguingly, altering this order as shown in the following expression may seem more natural:
\begin{align}
e^{T^{(N-1)}}e^{T^{(N)}}|\phi_0^{(N-1)}\rangle. \label{eq: ans_new}
\end{align}
This new sequence, compared to the original dCC ansatz (\ref{eq: ans2}), seemingly maintains the same configurations in its expansion$-$implying that no configurations are omitted. However, a detailed analysis reveals differences in the interaction contributions to these configurations between the two dCC ans\"{a}tze. Crucially, the new dCC ansatz (\ref{eq: ans_new}) lacks certain significant contributions that facilitate the connection between $e^{T^{(N-1)}(t)}$ and $e^{T^{(N)}}$. These missing contributions are essential for bridging the operators $e^{T^{(N-1)}(t)}$ and $e^{T^{(N)}}$, which remain isolated in the dCC ansatz (\ref{eq: ans_new}). To account for these crucial interactions, the original dCC ansatz (\ref{eq: ans2}) is necessary. Below, we elaborate on the differences in the expansions of the two dCC ans\"{a}tze when applied to the three-electron state of the three-site SIAM.}

\textcolor{black}{Using the spin-orbital labeling provided in the inset of Figure \ref{fig:Gfxn_D}, the the $N$- and $(N-1)$-electron reference states are defined as follows:
\begin{align}
|\phi_0^{(N)}\rangle = a_3^\dagger a_2^\dagger a_1^\dagger a_0^\dagger |0\rangle; ~~~~
|\phi_0^{(N-1)}\rangle = a_3^\dagger a_2^\dagger a_0^\dagger |0\rangle. \notag 
\end{align}
Correspondingly, the coupled cluster operators $T^{(N)}$ and $T^{(N-1)}$ can be explicitly expressed as
\begin{align}
T^{(N)} &= t^{(N)}_{(4,0)} a_4^\dagger a_0 + t^{(N)}_{(4,2)} a_4^\dagger a_2 + t^{(N)}_{(5,1)} a_5^\dagger a_1 + t^{(N)}_{(5,3)} a_5^\dagger a_3 \notag \\
&~~~~ + t^{(N)}_{(4,5,1,0)} a_4^\dagger a_5^\dagger a_1 a_0 + t^{(N)}_{(4,5,3,0)} a_4^\dagger a_5^\dagger a_3 a_0 \notag \\
&~~~~ + t^{(N)}_{(4,5,1,2)} a_4^\dagger a_5^\dagger a_1 a_2 + t^{(N)}_{(4,5,3,2)} a_4^\dagger a_5^\dagger a_3 a_2;
\end{align}
\begin{align}
T^{(N-1)} &= t^{(N-1)}_{(4,0)} a_4^\dagger a_0 + t^{(N-1)}_{(4,2)} a_4^\dagger a_2 + t^{(N-1)}_{(1,3)} a_1^\dagger a_3 + t^{(N-1)}_{(5,3)} a_5^\dagger a_3 \notag \\
&~~~~ + t^{(N-1)}_{(4,5,3,0)} a_4^\dagger a_5^\dagger a_3 a_0 + t^{(N-1)}_{(4,1,3,0)} a_4^\dagger a_1^\dagger a_3 a_0\notag \\
&~~~~ + t^{(N-1)}_{(4,5,3,2)} a_4^\dagger a_5^\dagger a_3 a_2 + t^{(N-1)}_{(4,1,3,2)} a_4^\dagger a_1^\dagger a_3 a_2.
\end{align}
Here, $t^{(N)}$'s and $t^{(N-1)}$'s represent scalar amplitudes. To roughly elucidate the difference between the ans\"{a}tze (\ref{eq: ans2}) and (\ref{eq: ans_new}), consider their truncated Taylor expansions:
\begin{align}
&e^{T^{(N)}}e^{T^{(N-1)}}|\phi_0^{(N-1)}\rangle \notag \\
&~~~~\approx (1 + T^{(N)})(1 + T^{(N-1)})|\phi_0^{(N-1)}\rangle \notag \\
&~~~~ = (1 + T^{(N)} + T^{(N-1)} + T^{(N)}T^{(N-1)} ) |\phi_0^{(N-1)}\rangle, \\
&e^{T^{(N-1)}}e^{T^{(N)}}|\phi_0^{(N-1)}\rangle \notag \\
&~~~~\approx (1 + T^{(N-1)})(1 + T^{(N)})|\phi_0^{(N-1)}\rangle \notag \\
&~~~~ = (1 + T^{(N-1)} + T^{(N)} + T^{(N-1)}T^{(N)} ) |\phi_0^{(N-1)}\rangle,
\end{align}
from which the comparison of ans\"{a}tze (\ref{eq: ans2}) and (\ref{eq: ans_new}) approximately boils down to the difference between $T^{(N)}T^{(N-1)} |\phi_0^{(N-1)}\rangle$ and $T^{(N-1)}T^{(N)} |\phi_0^{(N-1)}\rangle$. Through their explicit expansions:
\begin{widetext}
\begin{align}
T^{(N)}T^{(N-1)} |\phi_0^{(N-1)}\rangle 
&= \big( t^{(N)}_{(4,0)} a_4^\dagger a_0 + t^{(N)}_{(4,2)} a_4^\dagger a_2 \big) \big( t^{(N-1)}_{(1,3)}  a_1^\dagger a_3 + t^{(N-1)}_{(5,3)} a_5^\dagger a_3 \big) 
+ t^{(N)}_{(5,3)} a_5^\dagger a_3 \big( t^{(N-1)}_{(4,0)} a_4^\dagger a_0 + t^{(N-1)}_{(4,2)} a_4^\dagger a_2 \big) \notag \\
&~~~~\underline{+ t^{(N)}_{(5,1)} a_5^\dagger a_1 \big( t^{(N-1)}_{(1,3)} a_1^\dagger a_3 + t^{(N-1)}_{(4,1,3,0)} a_4^\dagger a_1^\dagger a_3 a_0 + t^{(N-1)}_{(4,1,3,2)}a_4^\dagger a_1^\dagger a_3 a_2 \big)} \notag \\
&~~~~\underline{+ \big( t^{(N)}_{(4,5,1,0)} a_4^\dagger a_5^\dagger a_1 a_0 + t^{(N)}_{(4,5,1,2)} a_4^\dagger a_5^\dagger a_1 a_2 \big) t^{(N-1)}_{(1,3)} a_1^\dagger a_3} ; \label{exp1_A1} \\
T^{(N-1)}T^{(N)} |\phi_0^{(N-1)}\rangle 
&= \big( t^{(N-1)}_{(4,0)} a_4^\dagger a_0 + t^{(N-1)}_{(4,2)} a_4^\dagger a_2 \big) t^{(N)}_{(5,3)} a_5^\dagger a_3 
+ \big( t^{(N-1)}_{(1,3)} a_1^\dagger a_3 + t^{(N-1)}_{(5,3)} a_5^\dagger a_3 \big) \big( t^{(N)}_{(4,0)} a_4^\dagger a_0 + t^{(N)}_{(4,2)} a_4^\dagger a_2 \big), \label{exp2_A1}
\end{align}
\end{widetext}
one can observe that the critical non-vanishing terms marked by the underline in the expansion (\ref{exp1_A1}) are absent in the expansion (\ref{exp2_A1}).}


\section{\textcolor{black}{Convergence performance of dCC ansatz (\ref{eq: ans2}) in comparison with the CC ansatz (\ref{eq: ans1}) towards the stationary $(N-1)$-electron CC energy}}\label{sec: appB}

\begin{figure}
    \centering
    \includegraphics[width=0.9\linewidth]{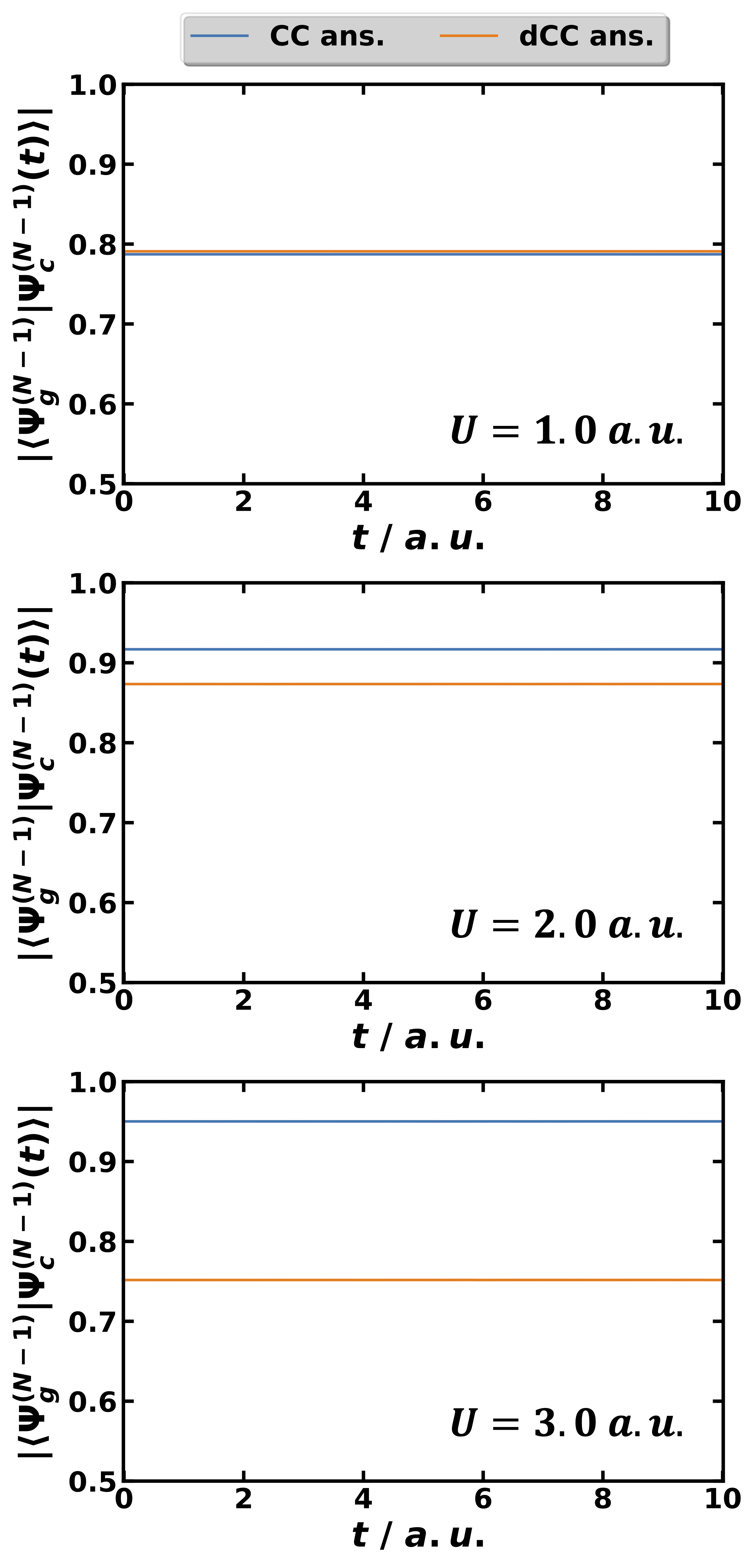}
    \caption{The overlap between the $(N-1)$-electron ground state and the time-dependent $(N-1)$-electron correlated state during the RT-EOM-CCSD simulations employing the CC ansatz (\ref{eq: ans1}) and the dCC ansatz (\ref{eq: ans2}). }
    \label{fig: ovlp_with_gs}
\end{figure}

\textcolor{black}{The time-averaged curves presented in Figure \ref{fig: En_SIAM}b,d exhibit a slightly different convergence performance between the ans\"{a}tze (\ref{eq: ans1}) and (\ref{eq: ans2}) towards the stationary $(N-1)$-electron CC energies. It is worth noting that, although the dCC ansatz (\ref{eq: ans2}) provides a more physically intuitive pathway$-$offering a potentially better-informed starting point for the $(N-1)$-particle correlated state, particularly in ionization scenarios where particle numbers change, the interaction between $T^{(N)}$ and $T^{(N-1)}$ introduces additional nonlinear behaviors. These behaviors could hinder the convergence of the average energy towards the stationary $(N-1)$-electron CC energies.
In essence, the ansatz that exhibits greater overlap with the dominant eigenstates of the Hamiltonian, typically low-energy states such as the ground state and low-lying excited states, will converge more rapidly in terms of time-averaged energies. For instance, for the three-site SIAM, Figure \ref{fig: ovlp_with_gs} compared the overlaps between the $(N-1)$-electron ground state of the Hamiltonian and the time-dependent $(N-1)$-electron correlated states derived from both CC and dCC ans\"{a}tze at different on-site interaction $U$'s. As can be seen, for $U=3.0$ a.u. the simpler CC ansatz (\ref{eq: ans1}) leads to a larger overlap with the ground state, implying a slightly faster convergence for the time-averaged CC energies. When $U$ decreases, the difference between the two ans\"{a}tze in terms of the overlap becomes smaller. }

\section{Connection to cumulant Green's function}\label{app:G_cumulant}

Eq. (\ref{eq: chgf1}) can be reformulated analogous to the cumulant Green's function formulation (see Appendix~\ref{app:G_cumulant}).
\begin{align}
    G_c^{\rm{R}}(t) &= -i \Theta(t) \exp(-i \epsilon_c t) \exp(C(t)) O^*(t)
\end{align}
where $\epsilon_c$ is the one-electron orbital energy of spin-orbital $c$ at, for example, the Hartree-Fock level, and the cumulant
\begin{align}
    C(t) &= \ln \big( \exp[-i (\Delta E_{CC}(t)-\epsilon_c) t] \big) \notag \\
    &= -i (\Delta E_{CC}(t)-\epsilon_c) t . \label{eq: cumulant}
\end{align}
From the Landau form of $C(t)$~\cite{1965417_Landau}
\begin{align}
    C(t) = \int \frac{\text{d}\omega}{2\pi} \frac{\beta(\omega)}{\omega^2}(\exp(i\omega t) - i\omega t -1)
\end{align}
and its second time-derivative
\begin{align}
    \frac{\partial^2 C(t)}{\partial t^2} = -\int \frac{\text{d}\omega}{2\pi} \beta(\omega) \exp(i\omega t) = -\beta(t),
\end{align}
the cumulant kernel $\beta(t)$ for Eq. (\ref{eq: cumulant}) can then be expressed as
\begin{align}
    \beta(t) = -i \frac{\partial}{\partial t} E_{CC}^{(N-1)}(t)  .
\end{align}
The cumulant kernel can also be used to compute the renormalization constant (a.k.a. the quasi-particle weight) $Z$ of the main peak in the computed spectral function through
\begin{align}
    Z = \exp(- a)~~\text{with}~~ a=\int \frac{\beta(\omega)}{\omega^2} \rm{d} \omega.
\end{align}
Note that a large $a$ suggests strong interaction effects, leading to a significant reduction in $Z$. Alternatively, $Z$ can also be identified from the Green's function or self-energy through
\begin{align}
    G(\omega) \approx \frac{Z}{\omega-\omega_0+i\eta},
\end{align}
or
\begin{align}
    Z = \left[ 1 - \frac{\partial \Re \Sigma(\omega) }{\partial \omega}\bigg\rvert_{\omega=\omega_0}\right]^{-1}.
\end{align}
Here we assume $G(\omega)$ as a pole at $\omega=\omega_0$, and $\eta$ is the broadening factor.

\section{Derive the components of $G_c^R$} \label{sec: decomp}

It is important to note that the inclusion of $\tilde{O}(t)$ in the formulation of $G_c^R(t)$, Eq. (\ref{eq: chgf2}), causes the Fourier transform of $G_c^R(t)$ to take a convolution form that depends on the Fourier transform of $\tilde{O}(t)$, $\tilde{\mathcal{O}}(\omega)$, and the Fourier transform of $A(t) := \tilde{N}_c(t)\exp(iE_{CC}^{(N)}t)$, $\mathcal{A}(\omega)$. 
For the latter, we have
\begin{align}
\mathcal{A}(\omega) &= \mathcal{F}\{\tilde{N}_c(t)\}(\omega) \bullet \delta(\omega - E_{CC}^{(N)})\notag \\
&= \int_{-\infty}^{\infty} \mathcal{F}\{\tilde{N}_c(t)\}(u) \delta(\omega - u - E_{CC}^{(N)}) \text{d} u \notag \\
&= \mathcal{F}\{\tilde{N}_c(t)\}(\omega - E_{CC}^{(N)}) \notag \\
&= \tilde{\mathcal{N}}_c(\omega - E_{CC}^{(N)}).
\end{align}
Here, $\delta(\cdot)$ denotes a Dirac delta function, and $\mathcal{F}\{\cdot\}$ denotes the Fourier transform. We are particularly interested in $\mathcal{F}\{\tilde{O}(t)\}$ and $\mathcal{F}\{\tilde{N}_c(t)\}$.

For the former, we have
\begin{align}
    \tilde{\mathcal{O}}(\omega) &= \left( \mathcal{F}\{\tilde{O}(t)\}(\omega) \right) \notag \\
    &= \langle \phi_0^{(N)} | (1+\Lambda^{(N)}) \overline{a_c^\dagger} \times \notag \\
    &~~~~ \mathcal{F}\{\exp(T^{(N-1)}(t))\}(\omega) | \phi_0^{(N-1)}\rangle. \label{eq: ovlp_fft}
\end{align}
Here, the expansion of $\exp(T^{(N-1)}(t))$ is given by
\begin{align}
    \exp(T^{(N-1)}(t)) &= \exp(\sum_{\mu} t_{\mu}^{(N-1)}(t) \mathbb{E}_{\mu}) \notag \\
    &= \prod_{\mu} \left(1 + t_{\mu}^{(N-1)}(t) \mathbb{E}_{\mu} \right) \notag \\
    &= 1 + \sum_{\mu} \tilde{t}_{\mu}^{(N-1)}(t) \mathbb{E}_{\mu} \label{eq: eT_expansion}
\end{align}
where $\mu$ is a compound index denoting the excitation, e.g. $\mu = (p,q)$ referring to a single excitation, $\mu = (p,q,r,s)$ referring to a double excitation, etc. $\mathbb{E}_{\mu}$ is the excitation generator, e.g. $\mathbb{E}_{(p,q)}=a_p^\dagger a_q$ and $\mathbb{E}_{(p,q,r,s)}=a_p^\dagger a_q^\dagger a_s a_r$. The relationship between $\tilde{t}_{\mu}^{(N-1)}(t)$ and $t_{\mu}^{(N-1)}(t)$ can be determined through cluster analysis of Eq. (\ref{eq: eT_expansion}). For example, if the CC operators only include the singles and doubles, we have:
\begin{align}
\tilde{t}^{(N-1)}_{(p,q)}(t) & = t^{(N-1)}_{(p,q)}(t)  ~, \\
\tilde{t}^{(N-1)}_{(p,q,r,s)}(t) & = t^{(N-1)}_{(p,q,r,s)}(t)+t^{(N-1)}_{(p,r)}(t) t^{(N-1)}_{(q,s)}(t)\notag \\
&~~~~-t^{(N-1)}_{(p,s)}(t) t^{(N-1)}_{(q,r)}(t). \label{eq:cluster_analysis}
\end{align}
Therefore, in Eq. (\ref{eq: ovlp_fft}):
\begin{align}
    \mathcal{F}\{\exp(T^{(N-1)}(t))\}(\omega) = \delta(\omega) + \sum_n \mathcal{F}\{ \tilde{t}^{(N-1)}_n (t)\}(\omega) \mathbb{E}_n. \label{eq: T_fft}
\end{align}
Regarding $\tilde{N}_c(t)$ and its Fourier transform, by noting that $[E^{(N-1)}_{dCC}]_t$ approaches the static CC energy of the $(N-1)$-particle system in the long $t$ limit (i.e., $\lim_{t\rightarrow \infty} [E^{(N-1)}_{dCC}]_t = E^{(N-1)}_{CC}$, see Figure \ref{fig: En_SIAM}b,d in the next section), we can write
\begin{align}
[E^{(N-1)}_{dCC}]_t  = E^{(N-1)}_{CC} + D(t)
\end{align}
where $D(t)$ is a damping function with $D(0)$ is a finite scalar and $\lim_{t\rightarrow \infty} D(t) = 0$. Therefore,
\begin{align}
\tilde{N}_c(t) &= \exp(-i E^{(N-1)}_{CC} t) \exp(-i D(t) t),
\end{align}
and from convoluted Fourier transform we have
\begin{align}
\tilde{\mathcal{N}}_c(\omega) &= \mathcal{F}\{ \exp(-i D(t) t) \} (\omega + E^{(N-1)}_{CC}),\\
\mathcal{A}(\omega) &= \mathcal{F}\{\exp(-i D(t) t)\}(\omega  + \Delta E_{CC}), \label{eq: FT_W}
\end{align}
where $\Delta E_{CC} = E^{(N-1)}_{CC} - E_{CC}^{(N)}$. 
Note that $\mathcal{F}\{A(t)\}(\omega)$ would contain a central peak at the frequency $\omega = \Delta E_{CC}$, corresponding to the long-time behavior of the function, i.e., a steady-state sinusoidal oscillation at that frequency. The initial transient behavior of $D(t)$ will influence $\mathcal{A}(\omega)$, but primarily outside the peak at $\omega = -\Delta E_{CC}$, causing additional spectral content such as spectral broadening and other peaks. 
In particular, the oscillatory behavior of $D(t)$ can introduce varying frequencies, which might manifest as additional peaks in $\mathcal{A}(\omega)$. These peaks will be at frequencies that correspond to significant components of $D(t)$'s oscillatory pattern. In this case, we can decompose $D(t)$ according to the cluster analysis of $[E^{(N-1)}_{dCC}]_t$ and $E^{(N-1)}_{CC}$, i.e.,
\begin{align}
[E^{(N-1)}_{dCC}]_t &= \langle \phi_0^{(N-1)} | \bar{H} ( 1 + \sum_n [\tilde{t}_n^{(N-1)}]_t \mathbb{E}_n ) | \phi_0^{(N-1)} \rangle, \\
E^{(N-1)}_{CC} &= \langle \phi_0^{(N-1)} | \bar{H} ( 1 + \sum_n \tilde{t}_n^{(N-1)} \mathbb{E}_n ) | \phi_0^{(N-1)} \rangle, \\
D(t) &= \sum_n d_n(t) , \label{eq: D_t}
\end{align}
where
\begin{align}
[\tilde{t}_n^{(N-1)}]_t &= \frac{1}{t}\sum_n \int_{t_0}^{t_0+t} \tilde{t}_n^{(N-1)}(\tau) {\rm d}\tau, \\
\Delta\tilde{t}_n^{(N-1)}(t) &= [\tilde{t}_n^{(N-1)}]_t - \tilde{t}_n^{(N-1)}, \\
d_n(t) &= \langle \phi_0^{(N-1)} | H \Delta\tilde{t}_n^{(N-1)}(t) \mathbb{E}_n ) | \phi_0^{(N-1)} \rangle \label{eq: d_n_t}
\end{align}
and $\tilde{t}_n^{(N-1)}$'s are obtained from the conventional CC calculation of the $(N-1)$-particle system. Therefore,
\begin{align}
&\mathcal{F}\{\exp(-i D(t) t)\}(\omega) \notag \\
&~~\approx \mathcal{F}\{\prod_{n=1}^s \exp(-i d_n(t) t)\}(\omega) \notag \\
&~~= \mathcal{F}\{\exp(-i d_1(t) t)\}(\omega) \bullet \mathcal{F}\{\exp(-i d_2(t) t)\}(\omega) \bullet \notag \\
&~~~~ \cdots \bullet \mathcal{F}\{\exp(-i d_s(t) t)\}(\omega). \label{eq: conv_Dt}
\end{align}
Here, we approximate $D(t)$ by its first $s$ significant components. From Eqs. (\ref{eq: ovlp_fft})$-$(\ref{eq:cluster_analysis}), (\ref{eq: FT_W}), and (\ref{eq: conv_Dt}), the component analysis of the $G_c^R(\omega)$ becomes straightforward. 

\section{Working expression of $G_c^R(\omega)$} \label{sec: G_workingEq}

To obtained an expression for the Green's function in the frequency domain, we can write Eq. (\ref{eq: FT_W}) as a sum of weighted, broadened poles, i.e.
\begin{align}
\mathcal{A}(\omega) &\approx \sum_{k=1}^{N_k} \frac{C_k}{i(\omega-\omega_k)+\eta_k} \label{eq: W_expansion}
\end{align}
with $\omega_1 = -\Delta E_{CC}$ corresponding to the main peak and $\omega_k$ ($k>1$) corresponding to the satellite peaks. $\eta_k$'s are the decay parameters influencing how broad each peak in the frequency domain becomes. The $C_k$'s are the normalization constants that ensure  $\sum_{k=1}^{N_k} C_k = 1/\pi$. 
From (\ref{eq: W_expansion}), $A(t)$ can be approximately rewritten as 
\begin{align}
A(t) &\approx \mathcal{F}^{-1} \left\{ \sum_{k=1}^{N_k} \frac{C_k}{i(\omega-\omega_k)+\eta_k} \right\} \notag \\
&= \sum_{k=1}^{N_k} C_k \exp((i\omega_k - \eta_k) t)~~  (t>0) \label{eq: W_fitting}
\end{align}
Note that $C_k$'s can be obtained by utilizing Eq. (\ref{eq: W_fitting}) to fit $A(t)$. 
Plugging (\ref{eq: W_expansion}) to the cumulant Green's function, and given
\begin{align}
\lim_{\eta_k\rightarrow 0^+} \frac{1}{i(\omega-\omega_k)+\eta_k} \approx \pi \delta(\omega-\omega_k),
\end{align}
we have the following approximation
\begin{align}
G_c^{\rm R}(\omega) &\approx -i \pi \sum_{k=1}^{N_k} C_k \tilde{\mathcal{O}}(\omega- \omega_k).\label{eq:G_comp}
\end{align}
For example, the leading terms in the expansion of (\ref{eq:G_comp}) can be elaborated as follows,
\begin{align}
G_c^{\rm R}(\omega) &\approx -i \pi \bigg( 
C_1 \tilde{\mathcal{O}}(\omega- \omega_1)  \notag \\
&~~~~~~~~~~~~~~~~ + \sum_{k=2}^{N_k} \frac{C_k}{i(\omega-\omega_k)+\eta_k} + \cdots
\bigg).
\end{align}
\bibliography{gfcc}
\end{document}